\documentclass[preprint]{aastex}

\newcommand{\etal}{{\em et~al.\,}}

\received{}
\revised{}
\accepted{}
\journalid{}
\articleid{}
\paperid{}
\ccc{}
\begin{document}

\title{CFHT Optical PDCS Survey II: Evolution in the Space Density of
Clusters of Galaxies\altaffilmark{1,2}}
\altaffiltext{1}{Based on observations collected at the
Canada France Hawaii Telescope operated by the National Research
Council of Canada, the Centre National de La Recherche Scientifique de
France, and the University of Hawaii}
\altaffiltext{2}{Based on observations
obtained with the Apache Point Observatory 3.5-meter telescope, which
is owned and operated by the Astrophysical Research Consortium}

\author{B. P. Holden\altaffilmark{3,4}}
\affil{Department of Astronomy and Astrophysics, University of
Chicago}
\affil{5640 South Ellis Ave. Chicago, Illinois 60637}
\email{holden@oddjob.uchicago.edu}
\altaffiltext{3}{Visiting Astronomer, Kitt Peak Observatory, National
Optical Astronomy Observatories, which is operated by the Association
of Universities for Research in Astronomy (AURA), Inc., under cooperative
agreement with the National Science Foundation.}
\altaffiltext{4}{Currently residing at: IGPP/UC-Davis, L-413, LLNL,
P.O. 808, Livermore, CA 94551}

\author{C. Adami} 
\affil{Department of Physics and Astronomy, Northwestern University,
2131 Sheridan Road, Evanston, Illinois 60208-2900 \\
IGRAP, Laboratoire d'Astronomie Spatiale, Traverse du Siphon, F-13012,
Marseille, France} 
\email{christophe.adami@astrsp-mrs.fr}

\author{R. C. Nichol}
\affil{Department of Physics, Carnegie Mellon University,
5000 Forbes Ave. Pittsburgh, Pennsylvania 15213-3890}
\email{nichol@andrew.cmu.edu}

\author{F. J. Castander}
\affil{Observatoire Midi-Pyrenees, 14, Avenue Edouard Belin,
31400 Toulouse, France \\
Department of Astronomy and Astrophysics, University of
Chicago, 5640 South Ellis Ave. Chicago, Illinois 60637}
\email{fjc@ast.obs-mip.fr}

\author{L. M. Lubin} 
\affil{California Institute of Technology, 105-24 Caltech, 1201 East
California Blvd, Pasadena, California 91125}
\email{lml@astro.caltech.edu}

\author{A. K. Romer\altaffilmark{3}}
\affil{Department of Physics, Carnegie Mellon University,
5000 Forbes Ave. Pittsburgh, Pennsylvania 15213-3890}
\email{romer@andrew.cmu.edu}

\author{A. Mazure} 
\affil{
IGRAP, Laboratoire d'Astronomie Spatiale, Traverse du Siphon, F-13012,
Marseille, France} 
\email{alain.mazure@astrsp-mrs.fr}

\author{M. Postman\altaffilmark{3,5}} 
\affil{Space Telescope Science Institute\altaffilmark{5}, 3700 San Martin Dr.,
Baltimore, Maryland 21218}  
\email{postman@stsci.edu}
\altaffiltext{5}{The Space Telescope Science Institute is operated by
the AURA, Inc., under National Aeronautics and Space Administration
(NASA) Contract NAS 5-26555.}
\and
\author{M. P. Ulmer} 
\affil{Department of Physics and Astronomy, Northwestern University,
2131 Sheridan Road, Evanston, Illinois 60208-2900} 
\email{m-ulmer2@nwu.edu}

\begin{abstract}

We present the first dynamical study of the optically selected Palomar
Distant Cluster Survey (PDCS).  We have measured redshifts for
seventeen clusters of galaxies in the PDCS and velocity dispersions
for a subset of eleven.  Using our new cluster redshifts, we
re-determine the X-ray luminosities and upper limits.  We show that
eleven out of twelve PDCS clusters we observed are real over-densities
of galaxies.  Most clusters have velocity dispersions appropriate for
clusters of galaxies.  However, we find a fraction ($\sim$
\slantfrac{1}{3}) of objects in the PDCS which have velocity
dispersions in the range of groups of galaxies (200 km s$^{-1}$ $\pm$
100 km s$^{-1}$) but have richnesses appropriate for clusters of
galaxies.

Within our survey volume of $31.7^{+0.5}_{-0.8}\times 10^4$ h$^{-3}$
Mpc$^3$ ($q_o = 0.1$) for Richness Class 2 and greater clusters, we
measure the richness function, X-ray luminosity function (using both
the detections and upper limits), and the mass function derived from
our velocity dispersions.  We confirm that the space density, as a
function of richness, of clusters of galaxies in the PDCS is $\sim$ 5
times that of the Abell catalog.  Excluding the above fraction of
\slantfrac{1}{3} of objects with low velocity dispersions, we measure
a space density $\sim$ 3 times that of the Abell catalog for
equivalent mass clusters of galaxies, raising the possibility that the
Abell catalog is incomplete.  However, our space density estimates are
in agreement with other low-redshift, optically-selected cluster
surveys such as the EDCC, APM and EDCC2.  Our X-ray luminosity
function agrees with other measurements based on both X-ray and
optically selected samples, so we find that the PDCS does not miss
clusters of galaxies that would be found in an X-ray selected survey.
Our resulting mass function, centered around $10^{14}$ M$_{\sun}$
$h^{-1}$, agrees with the expectations from such surveys as the
Canadian Network for Observational Cosmology cluster survey, though
errors on our mass measurements are too large to constrain
cosmological parameters.  We do show that future machine-based,
optically-selected surveys can be used to constrain cosmological
parameters.

\keywords{galaxies: clusters: general --- catalogs --- cosmology: observations}

\end{abstract}

\section{Introduction}

The space density of virialized clusters of galaxies as a function of
mass and redshift are primary predictions of theories on the formation
and evolution of structure.  In general, three parameters control the
shape, normalization and evolution of the mass function of clusters of
galaxies.  These parameters are the density of matter in the universe,
$\Omega_m$, the variance of the distribution of mass density
fluctuations at cluster scales, $\sigma_8$, and the shape of the
spectrum of density fluctuations, commonly quantified as $\Gamma$
\citep{press74,efstathiou92,lacey93,viana96, Eke96, Kitayama97}.
We can directly test these theoretical predictions and, therefore,
constrain these parameters, by measuring masses of clusters of
galaxies in a sample with a known survey volume.

Most efforts to measure the space density of clusters of galaxies as a
function of mass have focused on X-ray selected samples.  X-ray
selection has two advantages.  First, X-ray luminosity is strongly
correlated with mass.  Second, X-ray selected surveys have a selection
function that is easy to quantify.  Therefore, previous work focused
on the mass function derived from velocity dispersions of X-ray
selected clusters \citep[as examples]{carlberg97b, borgani99}, the
temperature function \citep[for an example]{henry97} or the luminosity
function \citep[as examples]{reichart99, borgani99b}.

There is a large spread in the above results.  Yet, most of the papers
mentioned above use the sample of clusters of galaxies from the
Extended Medium Sensitivity Survey or EMSS
\citep{gioia90b,gioia90a,henry92,nichol97}.  For example,
\citet{reichart99} finds the most likely value of $\Omega_m$ from the
mass function of the EMSS to be around 1, while the CNOC survey
\citep[for example]{carlberg96} finds a the most likely value to be to
0.2.  Recently, \citet{borgani99} finds that $\Omega_m$ from the CNOC
survey can be constrained to the range $0.35 < \Omega_m < 1.0$.
\citet{borgani99} finds that the large uncertainty in the best fitting
values of $\Omega_m$ is partly the result of the uncertainty in the
mass function of low redshift clusters of galaxies.  However, the high
redshift sample of the EMSS only contains the highest mass clusters
and, furthermore, is incomplete even at those masses (see Borgani
\etal 1999a\nocite{borgani99} for a discussion of the completeness of
the sample used by Carlberg \etal 1997\nocite{carlberg97b}, Borgani
\etal 1999a\nocite{borgani99} estimate that the CNOC sample contains
only 25\% of all clusters with $\sigma_v > 800$ km s$^{-1}$ because of
the X-ray threshold of $ >10^{45}$ erg s$^{-1}$).  Though there is
great potential in the CNOC sample, both a larger number of clusters
and a larger range in masses at both high and low redshift is needed
to constrain cosmological parameters.

To increase the range in mass at high redshift, we have used optically
selected clusters of galaxies.  Specifically, we have collected a
number of redshift measurements towards clusters of galaxies in the
Palomar Distant Cluster Survey \citep[PDCS]{postman96} so we can
measure velocity dispersions, the first such velocity dispersion
measurements of this catalog.  With our sample of $0.2 \le z \le 0.6$
clusters of galaxies, we could potentially improve the sample of
\citet{carlberg97b} and make a better measurement of the value of
$\Omega_m$.  The PDCS contains a much larger space density of clusters
of galaxies, between $\sim 10^{-5}$ and $\sim 10^{-6}$ h$^{-3}$
Mpc$^{3}$, in the same redshift range as the EMSS, and, therefore,
should contain lower mass clusters.  We should then have a sample
that, when combined with the EMSS, increases the dynamic range of the
cluster mass function making a more robust measurement of $\Omega_m$.

The approach of the PDCS was to create a model of what clusters of
galaxies look like, called a matched-filter.  The model for the galaxy
distribution is a Schechter function for the luminosity distribution
of the cluster galaxies and a profile of \(\frac{1}{\sqrt{1 +
(r/r_c)^2}} \).  The core radius of the radial profile were
fixed as were the slope and ``knee'' of the Schechter luminosity
function.  The total cluster size was also restricted to 1 $h^{-1}$
Mpc.  The strength of any observed correlation of the observed galaxy
catalog with this matched-filter can be used to measure how well the
model matches the data, with the strongest correlations being assigned
to cluster candidates.  In this way, the PDCS also generated an
estimated redshift, a galaxy richness based on the normalization of
the luminosity function and other parameters for each cluster
candidate based on the best-fitting model.  The advantage of this
approach has lead other groups to use similar techniques, see for
example \citet{dalton97}, \citet{kepner99}, \citet{olsen99} and
\citet{bramel99}.

Using only the derived quantities from the cluster catalog, the
authors of the PDCS found that the space density of clusters of
galaxies was $5\ \pm\ 2$ times that of the space density in the Abell
catalog (though the space density is consistent with that found in low
redshift automated catalogs such as the Edinburgh-Durham Cluster
Catalog and APM cluster catalogs).  Secondly, the authors found no
evidence for evolution in the space density with redshift.  Both of
these results rely on the cluster catalog alone, a catalog based
entirely on imaging data.  

We have completed a program of obtaining redshifts and velocity
dispersions of PDCS clusters to measure the space density as a
function of mass as well as richness and X-ray luminosity.  This will
allow us to both test the original results of the PDCS and possibly
provide a complementary sample to that of \citet{carlberg97b}.  We
began this program with the X-ray survey of \citet{holden97}, or H97.
Our first spectroscopic observations are described in
\citet{holden99}, H99, but the majority of spectra are from the CFHT
Optical PDCS survey which is described in \citet{adami99}.  We
summarized our data in \S 2, with an emphasis on what changes we have
made from the previously mentioned papers.  We then derive cluster
redshifts, velocity dispersions, X-ray luminosities and masses in \S
3.  Using the survey volume estimated in \S 4, we find the richness
and X-ray luminosity function of PDCS clusters in \S 5.  We compute
the mass function of PDCS clusters, computed in \S 6, and find it is
consistent with the mass function found for the sample of
\citet{carlberg97b} given specific choices for the relation between
$\sigma_8$ and $\Omega_m$, the value of $\Gamma$ and the form of the
richness-mass relation.  Finally, in \S 7 we summarize our results and
discuss the future prospects of using optically selected clusters of
galaxies to probe theoretical models of cluster formation and
evolution.  Unless otherwise noted, we use $q_o = 0.1$ ($\Omega_m =
0.2$ and $\Lambda = 0$) and $H_o = 100\ h\ {\rm km\ s^{-1}}$.

\section{Data}

Our study comprise of three data sets: the X-ray survey of the PDCS
from H97, the spectroscopic sample of H99 and the larger spectroscopic
sample of \cite{adami99}.  The spectroscopic survey from
H99\nocite{holden99} contains $\sim$ 100 spectra of galaxies towards
sixteen PDCS clusters.  The second set of galaxy spectra, the survey
described in \citet{adami99}, contains 636 redshifts towards eleven
PDCS clusters.  Below, we shall discuss the important details on how
each dataset was constructed.

\subsection{X-ray Imaging Data}

The original X-ray imaging data reduction is discussed in
H97\nocite{holden97}.  We have repeated the reduction using the new
cluster redshifts and in light of the experience of the Bright
Serendipitous High-redshift Archival ROSAT Cluster survey (SHARC;
\cite{romer99}).  We shall discuss below the process of image
preparation and measuring count-rates for the PDCS cluster candidates
with X-ray images.  We shall highlight where we used a different
approach than in H97\nocite{holden97}.

First we selected PDCS cluster candidates.  For a PDCS cluster
candidate to be included in our X-ray sample, we required that its
optical centroid be less than $40'$ from an X-ray image center.
Moreover, the exposure time at the optical centroid had to exceed 3000
seconds in the X-ray image. Thirty-one PDCS cluster candidates met
these requirements, see Table \ref{rawxraydata} for the list of
clusters and H97 for the X-ray data we used.

For each candidate, we derived an aperture for the flux measurement
using a ``beta'' model based on a modified isothermal sphere.  We used
values for the slope (\( \beta =\frac{2}{3} \)) and core radius (\(r_c
= 125\ h^{-1} \) kpc) which are typical for rich clusters
\citep{jonesforman92}.  We converted the above model from physical
units to angular units using the estimated redshifts from the PDCS
unless a spectroscopic redshift was available.  We then convolved the
model with empirical model point spread function (PSF) from
\citet{nichol94}.  We chose an aperture that contained 80\% of the
total flux of the PSF convolved model, an increase from the 70\% used
in H97\nocite{holden97}.  We increased the radius based on our
experience in the Bright SHARC survey \citep{romer99}, where the
larger aperture yielded more accurate flux measurement.  The resulting
aperture radii are listed in column five of Table \ref{rawxraydata}
and range from 2\farcm 0 to 5\farcm 0 .

Before measuring count-rates, we masked out certain pixels.  We masked
those pixels that were common to more than one cluster aperture and
those pixels that had less than 3000 seconds of exposure.  We also ran
the source detection algorithm from the Bright SHARC survey on the
pointings. We mask any pixels that contained flux from
detected sources with centroids more than 1.5\arcmin\ (three times the
uncertainty in the PDCS positions) from the PDCS cluster candidates.
We note that in two cases, PDCS 36 \& 62, we detected a source within
1.5\arcmin\ of the PDCS candidate and we assumed that these detections
correspond to those clusters.

After measuring the background subtracted count-rates, we determined
which of the thirty-one observations could be classified as
detections. For this, we used the (0.4 - 2.0 keV) error map from the
ESAS package.  The error map is simply the square root of the number
of counts in each pixel divided by the exposure time.  We converted
this back into a map of raw counts.  We used the ESAS error map
instead of the count-rate map because the ESAS software removes
various backgrounds from the count-rate map which are not removed from
the error map.  In H97\nocite{holden97} we assumed that a net
count-rate that was three times the error estimate represented a three
$\sigma$ (or a 99.87\% probable) detection.  As the error estimate is
derived by the square root of the number of counts in an aperture,
this is not correct for a small number of counts.  For our new
analysis of the X-ray data, we computed the counts in the aperture and
the counts expected in the aperture based on the counts in the
background annulus.  We then computed whether or not a fluctuation in
the background could have produce the counts in the aperture.  If the
counts in the aperture happened less than 0.13\% (equivalent to a
three $\sigma$ fluctuation for a Gaussian distribution) of the time by
chance, we considered that PDCS cluster to be detected. 

Seven PDCS cluster candidates met the detection criterion and they are
marked with an asterisk in Table \ref{rawxraydata}, column one.  Two
of the detections in Table \ref{rawxraydata} were upper limits in
H97\nocite{holden97}.  These are now considered detections because of
improvements in the source finding algorithm and the increase aperture
used to measure fluxes.  In Table \ref{rawxraydata}, we list the
background subtracted count-rate (column seven).  For the 24 PDCS
cluster candidates that were not detected, we list an upper limit to
their background subtracted count-rate.  Our the upper limits are the
count-rate that would produce enough counts to exceed the 99.87\%
confidence limit.

\subsection{Optical Spectra}

We have optical spectra from two samples.  The first sample was
collected with the Kitt Peak Mayall 4m and with the Astrophysical
Research Consortium 3.5m telescope.  This sample is more fully
described in H99\nocite{holden99}.  The second set of data were
collected using the Canada-France-Hawaii 3.5m Telescope and is
described in \citet{adami99}.  The differences
between these data sets are in the sample selection and in the number
of spectra taken per cluster candidate.  Below we will summarize the
candidate selection and the data reduction techniques used for both
data sets.

\subsubsection{KPNO and ARC Data}

Thirteen of the sixteen target clusters used in ARC/KPNO spectroscopic
survey were selected from the subsample discussed in \S 2.1.  The
remaining three (PDCS 11, 12 \& 23) were taken from the rest of the
PDCS.  For details on the spectral extraction and redshift measurement
for the ARC/KPNO sample, see H99.  

The main goals of the ARC/KPNO survey were to check the measured space
density of PDCS clusters and the accuracy of the matched-filter
estimated redshifts.  Therefore, we selected our targets as
independently as possible of the derived parameters given by PDCS.  We
selected targets using only the net number of $V_4\ <\ 21$, $V_4-I_4\
>\ 1$ galaxies (the subscript 4 refers to the 4-Shooter camera used to
construct the PDCS, see Postman \etal 1996\nocite{postman96} for
details on the filter system used and the resulting galaxy catalog)
within a 2\farcm 5 radius aperture of the PDCS cluster candidates'
position.  By taking this approach, we still select clusters that
potentially have a true redshift vastly different from the PDCS
estimated redshift.  

\subsubsection{CFHT Data}

Like the ARC/KPNO sample, the sample of PDCS cluster candidates
observed at the CFHT was based on those clusters in the X-ray sample
of H97\nocite{holden97}.  However, we selected clusters based on their
estimated redshifts and their richnesses.  We selected clusters
differently from the ARC/KPNO for this sample for two reasons.  First
off, we wanted to create a Richness Class limited sample to compare
with the ESO Nearby Abell Cluster sample \citep{katgert96,mazure96}.
Secondly, in H99\nocite{holden99}, we show that the PDCS estimated
redshifts are quite accurate for lower redshift clusters.  Therefore,
we can select targets using that estimate and be confident that we
have a sample of clusters of galaxies that span a limited redshift
range.

The selected clusters were all Richness Class 1 or greater ($\Lambda_V
> 40$) observed in H97\nocite{holden97} in either the $\alpha$ =
09$^{\rm h}$ or $\alpha$ = 13$^{\rm h}$ fields.  The redshift range
was $ 0.3 \le z_{estimated} \le 0.5$ for both fields.  Additionally,
one cluster at 00$^{\rm h}$ and one at 16$^{\rm h}$ were observed.
All of these clusters observed are listed in Table \ref{cfhtobs} along
with the $\Lambda_V$ richness and the estimated redshift of the
cluster from \citet{postman96}.  PDCS 32, 39 and 45
were not part of our main sample but were observed with masks used for
other PDCS clusters.

We used the Multi-Object Spectrograph (MOS) at the CFHT to observe
around 50 galaxies per mask in an effort to measure velocity
dispersions.  For details on the how we constructed multi-object masks
as well as how as details on our spectra extraction and redshift
measurements for the CFHT sample, see \citet{adami99}.  Briefly, our
primary selection consisted of galaxies at or brighter than $M_{\star}
= -21.1$ in the $V_4$ band, no color information was used.  We
optimally placed the slit-lets using a Minimal Spanning Tree over the
9\farcm 4 by 8\farcm 4 MOS field of view.  We then filled in the
remaining blank spaces of the mask with fainter galaxies.  We adjusted
our exposure times so that we would successfully measure 70\% of the
redshifts in our primary sample.  Overall, we averaged 66\% success
for our primary sample, success rates for all of our masks are listed in
Table 2 of \citet{adami99}.  We measured our redshifts using
cross-correlation \citep[RVSAO]{kurtz98} for 90\% of the galaxies.
For the remaining 10\%, we determined our redshifts using emission
lines only.

\section{Physical Quantities}

Here we discuss how we use the observed redshifts and X-ray
count-rates to estimate physically quantities.  We begin with how we
determined cluster redshifts and velocity dispersions.  Using those
redshift measurements, we can then compute X-ray luminosities.  With
the velocity dispersions, we can estimate the binding mass of the
clusters of galaxies in our sample.

\subsection{Determining cluster redshifts and velocity dispersions}

We have two different samples of PDCS cluster candidate redshifts
which contain a different number of redshifts.  Therefore, we have two
different approaches for measuring the redshifts of the PDCS clusters.
The large number of redshifts per cluster for some of the CFHT sample
also allow us to measure the velocity dispersions.  We compare our
technique for measuring cluster velocity dispersions with that used by
the Canadian Network for Observational Cosmology's (CNOC) survey of
clusters from the EMSS.  We then compare our measured redshifts with
the estimated redshifts of the PDCS cluster finding algorithm.

\subsubsection{ARC/KPNO cluster redshifts}

All PDCS cluster candidates in the ARC/KPNO survey had fewer than
eight redshifts per cluster, most had only four or five.  Therefore,
there was only one, if any, peak in the redshift distribution.  To
estimate the cluster redshift, we computed the median of all the
galaxy redshifts for that cluster.  If we had three or more concordant
cluster members within 1500 km s$^{-1}$ of the median redshift, we
considered the candidate to be a cluster of galaxies.  We also
calculated the mean redshift of all galaxies within 1500 km s$^{-1}$
of the median, in no case did the average redshift change by more than
400 km s$^{-1}$ from the median redshift of the whole sample.  The
median was a stable estimate of the cluster redshift because, as can
be seen in Table \ref{otherzs}, most if not all of the galaxies
observed were within the 1500 km s$^{-1}$ window on either side of the
median (PDCS 02 being the notable exception).  If the cluster was not
re-observed as part of the CFHT sample, we will use the median cluster
redshift, in Table \ref{otherzs}, for the rest of this work.  See H99
for further discussion on how determined cluster redshifts for the
ARC/KPNO sample.

\subsubsection{CFHT cluster redshifts and velocity dispersions}

In the CFHT sample, we had a much larger number of galaxy redshifts in
direction of our PDCS clusters.  Simply taking the median of all the
available redshifts would not be sufficient for two reasons. First,
the large number of intervening field galaxies would eliminate the
signal from the cluster and, second, we selected our fields to cover
multiple PDCS clusters so there are multiple peaks in the redshift
distribution.  Instead, we implemented a two step process.  The first
step is based on the process outlined in \citet{katgert96}.  At each
galaxy redshift we searched to see if five or more galaxies were
separated by less than 1000 km s$^{-1}$.  For each such over-density,
we then found the center and width of the group using the biweight
estimators of \citet{beers90}.  The results for the PDCS clusters in
our sample are listed in Table \ref{zmeans}.  The only PDCS cluster
not found with this technique was PDCS 34.  However, with the addition
of the redshifts from the ARC/KPNO survey, PDCS 34 becomes a clear
over-density in velocity space.

To assign an over-density in redshift to a PDCS cluster, we calculated
the mean Right Ascension and Declination of all of the galaxies within
three times the biweight width of the biweight center.  Then, we found
the PDCS cluster with the closest position.  In some cases we had to
make large offsets ($\sim$ 1\arcmin) between the mean position based on
the redshifts and the PDCS position.  We found this resulted from a
galaxy or galaxies far from coordinates the PDCS center but was in the
redshift range of the cluster.

Once we found over-densities in the galaxy redshift distribution, each
over-density was examined individually.  We used a maximum likelihood
algorithm to fit a single Gaussian and a flat background to the galaxy
redshifts using the technique outlined in \citet{cash79}.  For the
range in redshift over which to fit the Gaussian, we chose five times
the biweight scale determined above.  If the biweight was smaller than
the resulting fit, we increased the redshift to include five times the
fitted Gaussian sigma.  From the fitted Gaussian, we determine the
velocity dispersion of the clusters as well as the mean of the cluster
redshifts.

\begin{figure}
\begin{center}
\includegraphics[height=7in]{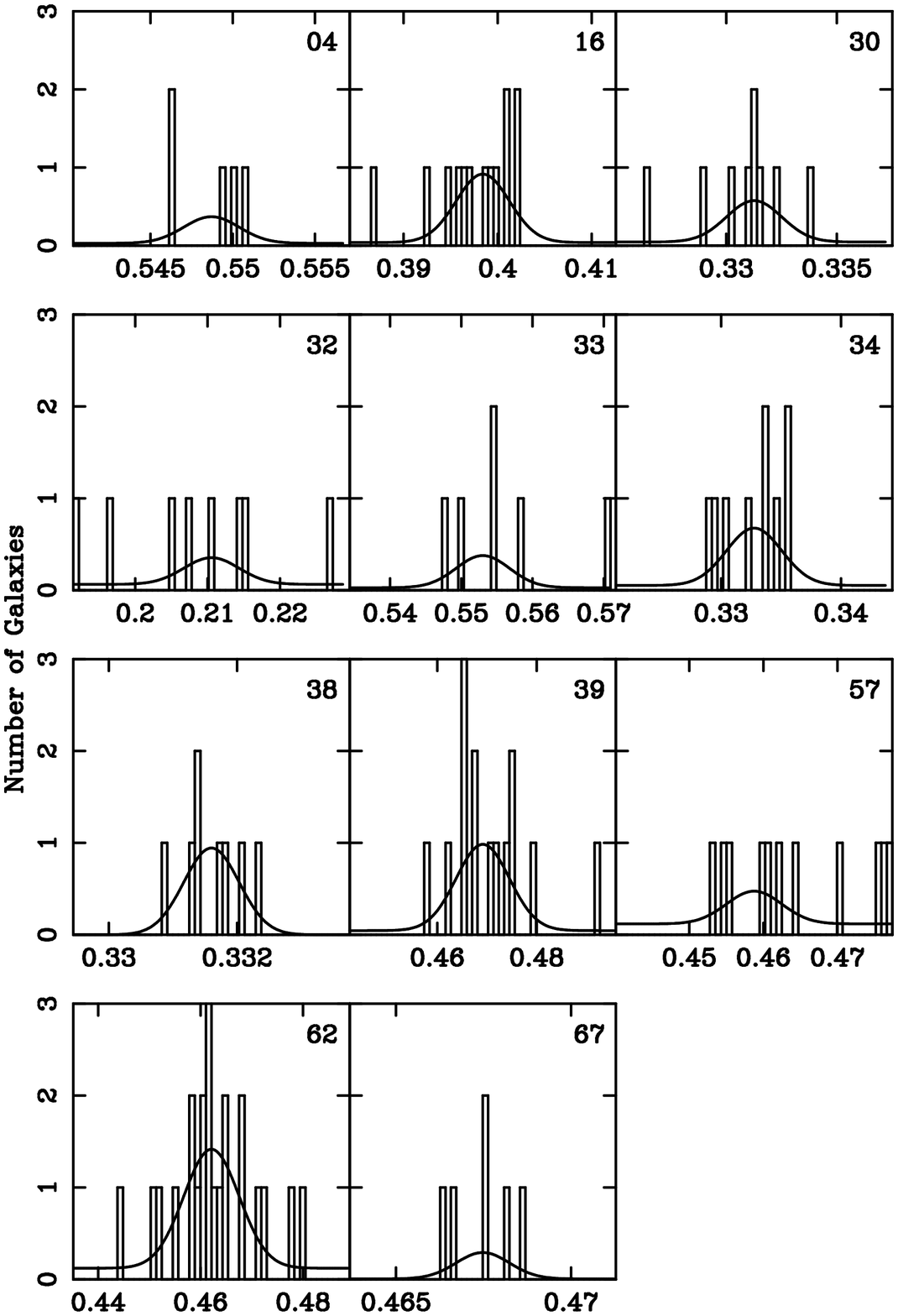}
\end{center}
\caption[holden.fig1.ps]{
Redshift distributions of PDCS clusters in CFHT sample.  The
histograms contain all redshifts within 5$\sigma_v$ of the mean of the
CFHT data.  The solid line is the Gaussian plus fit to the redshift
data. \label{my_hist}}
\end{figure}

In Table \ref{zmeans}, we show our resulting cluster redshifts and
velocity dispersions with errors.  We plot these fitted Gaussians in
Figure \ref{my_hist} on top of our redshift data.  We derived the 68\%
confidence limits in Table \ref{zmeans} using the method outlined in
\citet{cash79}.  The error estimates rely on assuming that
the likelihood distribution follows a $\chi^2$ distribution.  This
assumption is true in the case of a large number of data points.  For
many of our clusters, we do not have a large number of cluster
redshifts.  This means our error distribution is probably not
Gaussian, at least not for large confidence limits ({\em i.e.} $3
\times \sigma \ne 99.87$\% confidence limit).

Our maximum likelihood velocity dispersions are smaller than the
values derived using the biweight scale from \citet{beers90}.  The
median ratio of the biweight scale to the maximum likelihood standard
deviation is $1.18 \pm 0.17$.  \citet{carlberg97} finds a ratio of
$1.05 \pm 0.04$ between the biweight scale and their velocity
dispersion estimate, a smaller ratio than ours but not statistically
different.  Inter-loping field galaxies generally make it much easier
to over-estimate then under-estimate the velocity dispersion of a
cluster of galaxies, even when using a robust estimator such as the
biweight scale.  Our maximum likelihood approach explicitly accounts
for the background galaxy distribution, so we shall use those
estimates for the rest of this work.

Two clusters observed in the CFHT sample, PDCS 45 \& 61, are not
listed in Table \ref{zmeans} because the Gaussian fitting routine did
not converge.  For the rest of this work, we will not consider these
objects as confirmed clusters of galaxies.  However, neither of the
these two are part of our $\Lambda_V \ge 40$ sample, so our space
density estimates are not affected.  We would like to note that the
total integration time for PDCS 61 was quite low (20 minutes, see
\cite{adami99}), so it is unlikely we would have detected a cluster
of galaxies.  

\subsubsection{Comparison with the Results from the CNOC survey}

To check our technique of estimating velocity dispersions, we measured
the velocity dispersions of the eight $z \ge 0.3$ clusters \citep{
ellingson97, abraham98, ellingson98, yee98} in the CNOC survey.  We
can then compare our estimates with those of the CNOC survey.

The CNOC collaboration used a similar approach for determining the
cluster velocity dispersions.  The authors of the CNOC survey assumed
a flat field galaxy background within 15 $\sigma$ of the cluster and
used an iterative approach to determine the range in redshift over
which to measure the cluster velocity dispersion, see
\citet{carlberg96}.  However, the authors used the standard deviation 
as estimate of the velocity dispersion, instead fitting a Gaussian as
we did and the range in redshift used to measure the standard
deviation was chosen by visual inspection. 

\begin{figure}
\begin{center}
\includegraphics[height=5in]{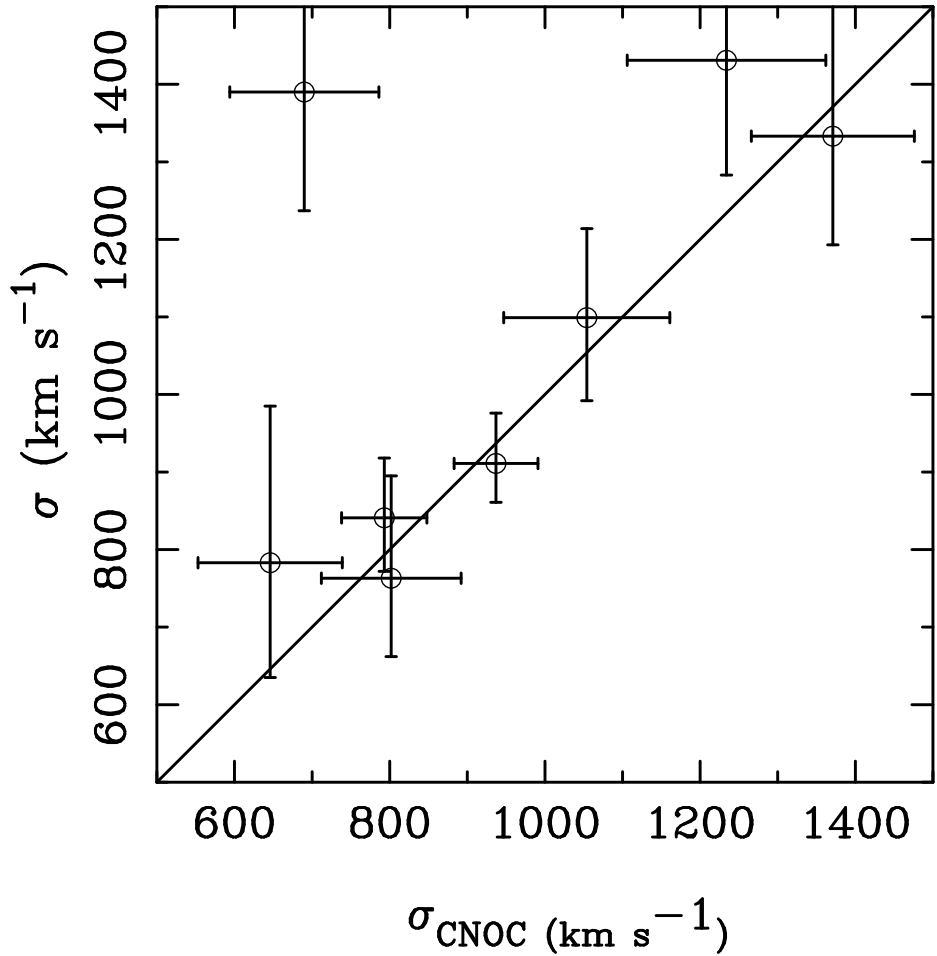}
\end{center}
\caption[holden.fig2.ps]{Comparison of velocity dispersion
estimates for the CNOC clusters.  The x-axis are those estimates from
Carlberg \etal (1996) while the y-axis are the
estimates based on the maximum likelihood technique discussed in this
work.  We also plot a 45\arcdeg\ line.  The point in strong
disagreement with the 45\arcdeg\ line is MS1512+36.\label{cnoc_compare}}
\end{figure}

\begin{figure}
\begin{center}
\includegraphics[height=5in]{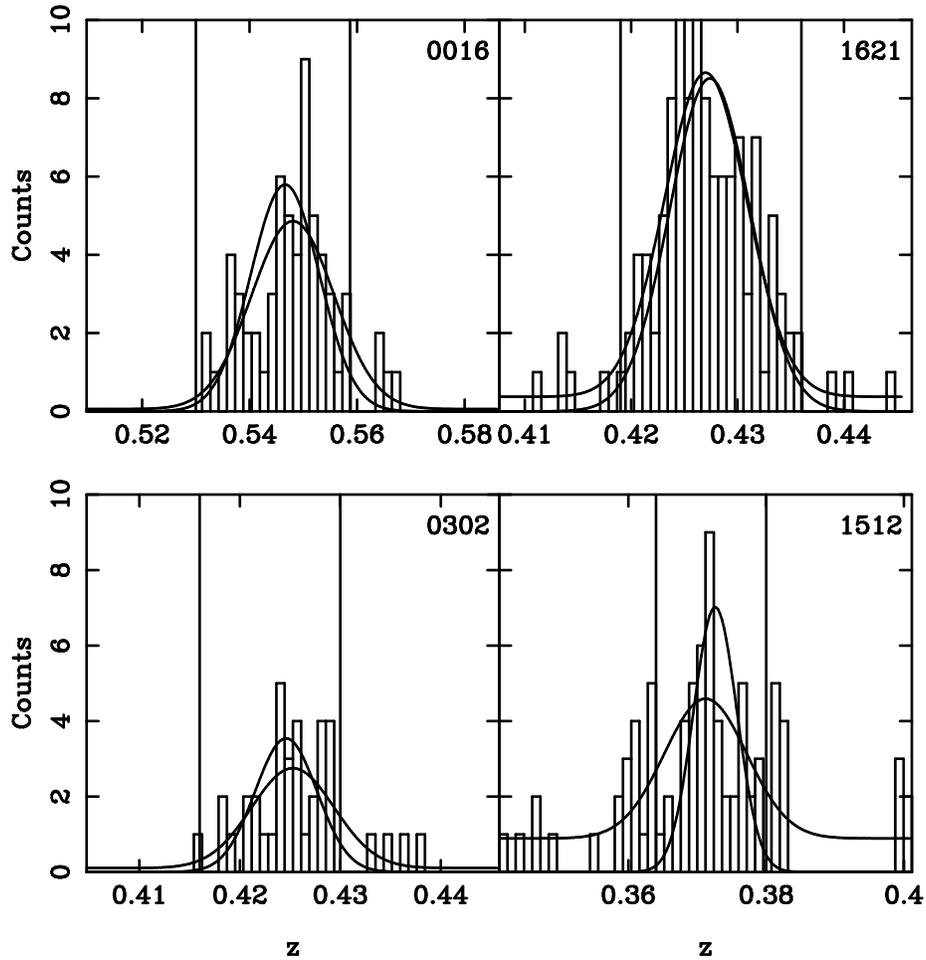}
\end{center}
\caption[holden.fig3.ps]{Redshift distributions of four CNOC clusters
with velocity dispersions.  The histograms contain all redshifts
within 5$\sigma_v$ of the mean.  The solid line with a background is the
estimate from above. The solid line originating from the x-axis is the
estimate from \cite{carlberg96}.  The two solid vertical lines
represent the range in redshift used by \cite{carlberg96} for
estimating the velocity dispersion.  The bottom two histograms are of
the two clusters that our estimate has the largest deviation from the
CNOC estimate.\label{cnoc_hist}}
\end{figure}

In Figure \ref{cnoc_compare} we show our measurements of the
rest-frame velocity dispersion versus those published by the CNOC
survey.  Though most of our values agree within two standard
deviations, there is one exception in Figure \ref{cnoc_compare}.  This
exception is MS 1512+36, which has a number of peaks in the velocity
distribution \citep{ borgani99}, see Figure \ref{cnoc_hist}.  It is
likely that the much lower velocity dispersion from the CNOC survey is
correct as it agrees with the X-ray luminosity and X-ray temperature
of the cluster.  This illustrates that our technique works well for
most clusters, but not for dynamically complicated ones such as MS
1512+36 because of our explicit assumption of single Gaussian
distribution.  

Removing MS 1512+36, we find that the average ratio of our dispersion
measurements to those from the CNOC survey is $1.05 \pm 0.10$.  We
conclude our technique is equivalent to that of the CNOC for clusters with
simple velocity distributions.  Therefore, we can compare the
distribution of velocity dispersions in our sample with that of the
CNOC survey.

\subsubsection{Comparison with the Estimated Redshifts in the PDCS}

\begin{figure}
\begin{center}
\includegraphics[height=5in]{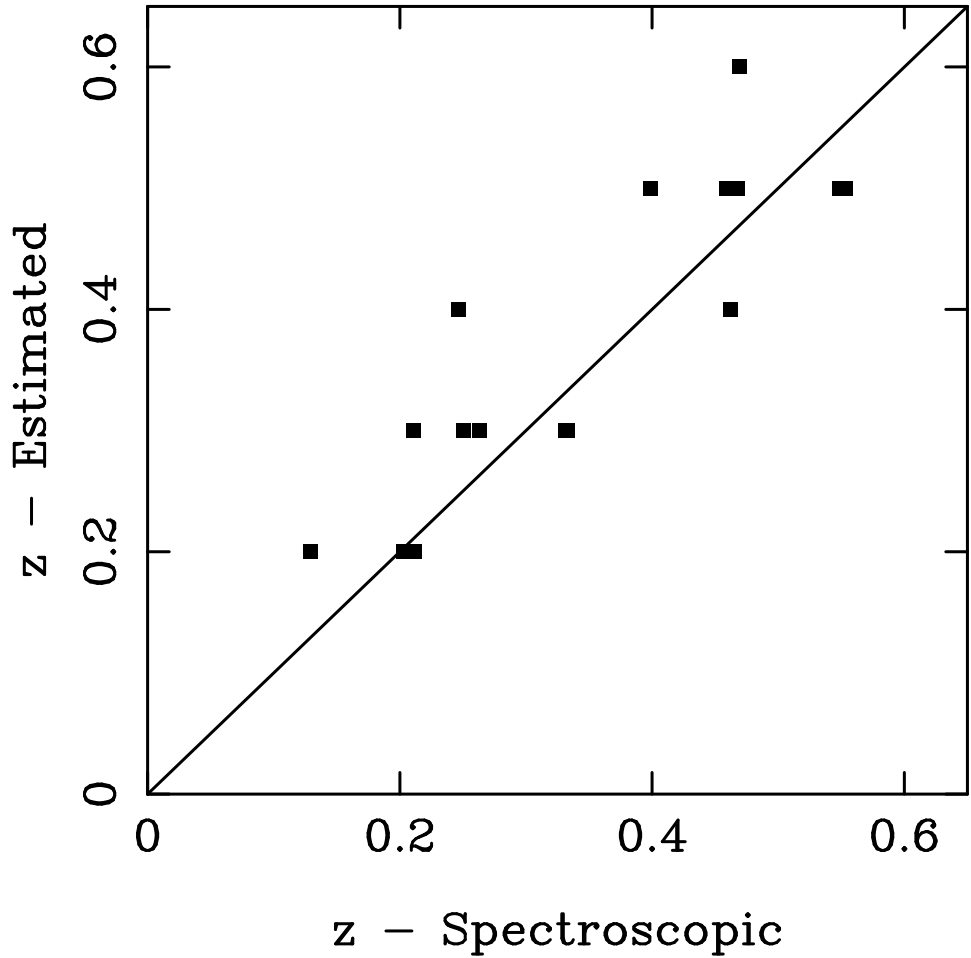}
\end{center}
\caption[holden.fig4.ps]{Measured redshifts compared match-filter estimated
redshifts.  We plot the PDCS estimated redshifts as a function of
spectroscopically measured redshifts for both the CFHT and the
ARC/KPNO sample.  We also plot a 45\arcdeg\ line, not a fit, for comparison. 
\label{zs_v_ph}} 
\end{figure}

One of the results of the PDCS cluster finding algorithm is an
estimated redshift.  In Figure \ref{zs_v_ph}, we plot the
estimated redshift versus our spectroscopically measured redshift as
well as the expected 45\arcdeg\ line for comparison.   

To quantify the observed accuracy of the matched-filter redshifts, we
calculated the standard deviation of the difference between the
matched-filter redshift and the spectroscopically measured redshift.
If we include only those clusters with three or more concordant galaxy
redshifts, as discussed above, we estimate the standard deviation to
be $\sigma_z = 0.063$ for seventeen clusters with $0.2 \le z_{est} \le
0.6$, almost exactly the same as the value we found in
H99\nocite{holden99}.  

As the scatter in the estimated redshift is important for modeling our
cluster selection procedure (see \S 4.3), we investigated the
distribution of the differences between the estimated redshift and the
spectroscopic redshift.  The average difference between these two
redshifts is 0.025, which we find to be not statistically significant
from zero according to a Student's t-test.  We also performed a
Kolmogorov-Smirnov test to compare the distribution of redshift
differences with a Gaussian with a zero mean and a variance as given
above.  We find no statistically significant difference between our
model and the data.  

Finally, we checked for any gross change in the distribution with
redshift. First we fit a straight line to the data in Figure
\ref{zs_v_ph}.  We then used a $\chi^2$ test and found no statistical
difference between the fitted line and a line with a 45\arcdeg\ slope
and no y-intercept (the line plotted in Figure \ref{zs_v_ph}).  Second
we divide the sample into two equal sized subsamples at $z_{estimated}
= 0.4$.  Using an F-ratio test, we find no difference between the
$z_{estimated} < 0.4$ sample of eight clusters and the $z_{estimated}
\ge 0.4$ sample of seven clusters.  Therefore, we do not find that the
apparent systematic shift in Figure \ref{zs_v_ph} is statistically
significant.

For the rest of this work, we will therefore assume the distribution
of real redshifts around an estimated redshift is a Gaussian with a
mean of zero and a standard deviation of $0.063^{+0.012}_{-0.010}$
(68\% confidence limits from an F-ratio test).

\subsection{New X-ray Energy Fluxes and Luminosities}

We converted our newly re-measured X-ray count-rates or upper limits
to energy fluxes by integrating a redshifted 1.0 keV thermal {\it
bremsstrahlung} spectrum over our energy passband of 0.4 - 2.0 keV.
We chose 1 keV, as opposed to a 6 keV spectrum as originally used in
H97\nocite{holden97}, since it is more representative of clusters in
our luminosity range \citep{reichart99b}.  All fluxes were then
corrected for absorption using the observed amount of Galactic neutral
hydrogen in the AT\&T Bell Laboratories 21 cm survey \citep{stark} and
the cross-section values from \citet{mm83}.  Once we had the energy
flux inside the masked aperture (column two of Table
\ref{xrayresults}), we needed to convert to a total flux for each
cluster candidate.  For cluster apertures where no part of the
aperture was masked, the energy flux was simply divided by 0.8 (see
previous discussion) to give a total flux (column four of Table
\ref{xrayresults}).  For the apertures that were masked, we computed
the fraction of the flux from our model that would fall within the
masked aperture and then used this fraction to convert between
measured and total flux.  The total flux (either upper limit or
detection) is listed in column three in units of $10^{-14}$ erg
s$^{-1}$ cm$^{-2}$ and the total luminosities are presented in units
of $10^{43}$ erg\ s$^{-1}$ in are listed in column four of Table
\ref{xrayresults}.  For the detections in Table \ref{xrayresults}, we
list errors on the fluxes and luminosities.  These errors are the 68\%
confidence limits for a Poisson distribution using the number of
counts in the PDCS cluster candidate aperture.  For upper limits, we
list the flux and luminosity equivalent to $3\sigma$ upper limit for a
Poisson distribution.

\subsection{Mass Estimates}

The simplest way to estimate the mass for a gravitationally bound
object in equilibrium is by using the virial theorem. This
requires an unbiased estimate of the velocity dispersion, $\sigma$ and
of the size, $r_v$, of the cluster of galaxies.

It has been shown in a number of papers, see for example in
\citet{crone94}, that the true virialized mass lies within a radius
defined by the ratio of the average density inside that radius to the
critical density or \( \Delta_{c}\).  Inside the radius where this
ratio is $\ge 200$\footnote{It was pointed out to us by Steve Kent
that there is a simple derivation for this.  The radius for which
crossing time for a cluster member equals the Hubble time has a mean
interior over-density of $\sim 200$.}, objects are found to be
virialized while outside this radius the process of accretion is
usually still occurring.  This motivated \citet{carlberg96} to define
a mass, $M_{200}$, based on the mass derived from the virial theorem
and combined with an isothermal sphere as a model of the cluster
potential.  This mass is completely specified by the velocity
dispersion \citep{bt87}.

A modification of the $M_{200}$ comes from \citet{borgani99}
\begin{equation} 
M(\sigma_v) = \left( \frac{\sigma_v}{1129\
f_{\sigma_v}\ {\rm km\ s^{-1}}} \right)^3 \left( \frac{\Delta_{c}}{178}
\right)^{-1/2} E^{-1}(z)\ 10^{15}\ h^{-1}\ M_{\odot} 
\label{msigma}
\end{equation} 
In that paper, they argue that $f_{\sigma_v} = 0.93$ and use \(
\Delta_{c} = 178 \) for $\Omega_m = 1$ instead of 200 as in
\citet{carlberg96}.  The term denoted $E(z)$ is proportional to the
time derivative of the logarithm of the scale factor, {\em i.e.} $H(z)
= E(z)^{1/2}\ H_o$ \citep{hogg99}.  These masses are $\sim 2/3$ of the
$M_{200}$ masses because the different constants used but otherwise
directly proportional.  For the rest of this work, we will use
equation \ref{msigma} to compute our mass estimates.

\subsection{Mass compared to richness}

In Figure \ref{richvmass}, we plot our masses as measured using
equation (\ref{msigma}) versus the PDCS V and I band richnesses.  We
use the mass estimate from equation (\ref{msigma}) as that is the
closest to mass estimator used in \citet{girardi98b}.  We also plot
the richness-mass relation from \citet{girardi98b}.  To convert
between the richness of the PDCS, $\Lambda$, and the Abell Richness
Class, we use the conversion from \citet{postman96} which we list
below in Table \ref{richdens}.

The comparison has quite a lot of scatter, even when considering the
large errors on our mass estimates.  Most of the PDCS clusters appear
to match the low-redshift richness-mass relation.  However, about
\slantfrac{1}{3} of the PDCS clusters fall below the expected line.
Many of these are low mass systems which have only a few galaxy
redshifts within 3$\sigma_v$ of the mean.  For example, PDCS 04 is a
Richness Class 3 cluster with only five galaxies within 3$\sigma_v$ of
the mean.  Using a simple $\chi^2$ statistic, we tested the idea that
our sample contains clusters whose masses were far to low for their
richnesses.  We find that we can reject at the 99.5\% confidence limit
the hypothesis that the PDCS mass-richness relation is the same as the
Abell mass-richness relation.  We find it likely, therefore, that some
of the clusters in the PDCS are true over-densities, but have their
richnesses over-estimated.

We fit a line to the richness-mass relation of \citet{girardi98b}
which has the following form
\begin{equation}
\log M (10^{14}\   h^{-1}\ M_{\odot})=  (1.34 \pm 0.26)\log \Lambda - 1.95 \pm 0.46 
\label{richnessmass}
\end{equation}
and we plot this relation in Figure \ref{richvmass}.  The fit is
rather uncertain because of the small number of data points and the
large scatter in mass at each Richness Class.  We note here that we
fit not the data from \citet{girardi98b}, but instead to the tabulated
richness-mass relation.  This is equivalent to fitting to the data
directly as the Abell Richness estimates are by class.  The tabulated
values are the solid circles in Figure \ref{richvmass}.

\section{Survey Volume}

An important part of any survey used for statistical purposes is the
estimate of the effective survey volume occupied by the objects in the
survey.  Below we will discuss our modeling of the selection criteria
and compute the resulting area and volume coverage for the three
surveys.

\subsection{X-ray Areal Coverage}

Our X-ray data covered parts of three of the five square-degree survey
areas in the PDCS.  In H97\nocite{holden97}, we calculated the overlap
between the PDCS survey areas and the ROSAT pointings to be 1.85
deg$^{2}$.  We computed these areas in H97\nocite{holden97} by finding all
pixels with exposure times greater than 3000 seconds and within
40\farcm 0 of the pointing centers.  We then found the
intersection of these pixels with the part of the sky covered by the
PDCS survey fields.  

The procedure we used in H97\nocite{holden97} underestimated the total area
covered in the X-ray data.  If an X-ray detection lay within 1\farcm 5
of a PDCS cluster candidate position, we considered that detection to
be associated with the PDCS cluster candidate.  Therefore, we extended
the length of the sides of the PDCS survey areas by 1\farcm 5 and
recalculated the overlap between the X-ray data and the PDCS survey
fields.  This procedure yielded an overlap 0.51 deg$^2$ for the
00$^{\rm h}$ field, 1.14 deg$^2$ for the 09$^{\rm h}$ field and 0.38
deg$^2$ for the 13$^{\rm h}$ field.  We find, with this improved
estimate, that the total area of overlap is 2.03 deg$^2$.  Our
CFHT sample covered 1.52 deg$^2$ of this area.

\subsection{Survey volume for the ARC/KPNO survey}

Our ARC/KPNO survey was a sub-sample from the whole of the X-ray
cluster candidate sample; {\em i.e.} 2.03 deg$^{2}$.  When determining
the survey volume, we must address two issues.  First, what is the
probability of a cluster at a given redshift to be selected for our
ARC/KPNO survey (hereafter $p_s(z)$)?  Second, once a cluster is
selected, how often would we successfully measure the redshift of that
cluster given our observing strategy ($p_r(z)$)?  These two
probabilities are then included in the following equation
\begin{equation} V = \int_{z_1}^{z_2} \frac{dV}{dz} p_s(z) p_r(z) dz
\label{surveyvol}
\end{equation}
where $\frac{dV}{dz}$ is determined by $q_o$, $H_o$ and
the survey area while the two probabilities, $p_s(z)$ and $p_r(z)$ are
determined by our selection criteria, the underlying cosmological
parameters and our observing strategy.  We performed Monte Carlo
simulations of both our selection procedure and our observing strategy
to answer these questions.  The details of these simulations are found
in H99\nocite{holden99} but below we will review the salient points.

To compute the probability of a cluster being selected for our survey,
$p_s(z)$, we constructed model PDCS clusters and simulated the
selection process outlined in \S 2.2.1.  We created model clusters
using a power-law profile \citep{lubin96} and a Schechter luminosity
function from \citet{Colless89} to represent the distribution of
positions and luminosities for the galaxies in the cluster.  For each
simulated cluster, we used random values for the parameters based on
the ranges specified in \citet{lubin96} and \citet{Colless89}.  The
normalization of the luminosity function was a randomly chosen
richness from the values in the PDCS catalog.  The resulting
probabilities from these simulations are illustrated, as a function
of richness and redshift, in Figure 4 of H99\nocite{holden99}.

To model our redshift completeness, which determines $p_r(z)$ in
equation (\ref{surveyvol}), we created artificial redshift distributions
using a Gaussian to represent the cluster velocity distribution and
the Canada-France Redshift Survey to represent the background
galaxy redshift distribution.  We used the same parameters and
observational limits as in H99\nocite{holden99} for determining the
number of cluster members inside the Gaussian velocity distribution.
Our simulation results can be seen in Figure \ref{probrvz} where we
plot the probability of successfully measuring the redshift of a
cluster in our ARC/KPNO sample.

To perform the integral in equation (\ref{surveyvol}), we must set the
bounds of the integral.  We chose $z_1 = 0.1$ as that is the lower
redshift limit of the PDCS.  For $z_2$, we computed where $p_r(z)$
dropped below 5\% which yields $z_2 = 0.55$ for Richness Classes 1 and
2 and $z_2 = 0.57$ for Richness Class 3.  These numbers are higher
than those found in H99\nocite{holden99}, because for that work we were
limited by a small number of redshift per cluster.  The addition of
the redshifts from the CFHT survey of the clusters in common with the
ARC/KPNO survey allow us to extend the redshift range of the ARC/KPNO
sample.  By extending the redshift range, we survey a larger volume
than in H99\nocite{holden99} and include more PDCS clusters.

For both sets of simulations, the important independent variables were
redshift and Richness Class.  The authors of the original PDCS survey
provide no results for Richness Class 0 clusters, so though we can
calculate a volume for those clusters, we will not consider those
clusters for statistical tests.  When we combine the results of both
sets of simulations, we found we surveyed $15.7\times 10^4$ h$^{-3}$
Mpc$^3$ for the Richness Class 1 clusters and $20.4\times 10^4$
h$^{-3}$ Mpc$^3$ for the Richness Class 2 clusters.

We investigated the change in enclosed volume as a function of the
various model parameters.  We assumed for all of these tests that
$p_r(z) = 1$, so we did not have to run two sets of simulations.  In
Table \ref{modelparams}, we show how changing various
parameters changes the volume surveyed for Richness Class 2 clusters.
The most surprising result is that by changing the absolute magnitude
of $M_{\star}$ by -0.1 magnitudes, an increase in the luminosity, the
change in the enclosed volume is only +0.2\% while a increase of 0.1
magnitudes decreases the volume by 10\%.  This small change in volume
for an increase in $M_{\star}$ stems from the steeply increasing
K corrections.  In the redshift range of $0.3 \le z \le 0.4$, the 4000
\AA\ break moves through the V filter causing the K corrections to
increase rapidly. For example, from z=0.36 to z=0.40, the K correction
changes by 0.2 magnitudes.  Therefore, an increase in $M_{\star}$ does
not increase the effective limiting redshift of the sample.

As a check of the K corrections, we tried the elliptical galaxy
K corrections, without the additional evolutionary corrections, of
\citet{poggianti97}, which we find changes the enclosed volume by
-10\%.  We can conclude, using the results of Table \ref{modelparams},
that the uncertainty in our survey volume from systematic shifts in
the individual model parameters is on the order of $\sim$10\%.

\subsection{Survey volume for the CFHT survey}

The PDCS cluster candidates in the CFHT survey were selected to be
within the estimated redshift range $0.3 \le z_{estimated} \le 0.5$
and have V-band richness $\Lambda_V \ge 40$ from the PDCS catalog in
\citet{postman96}.  We chose candidates for our
sample from the 09$^{\rm h}$ and 13$^{\rm h}$ fields as covered by the
X-ray survey from H97\nocite{holden97}. We observed all candidates that met
our criteria but PDCS 36.  The area we covered is 1.52 deg$^{2}$ with
the majority of that area in the 09$^{\rm h}$ field.

To compute the volume we surveyed, we need to compute $p_s(z)$ and
$p_r(z)$ for equation (\ref{surveyvol}), as we did for the ARC/KPNO
sample.  Instead of using Monte Carlo simulations as above, we chose
an analytical model for $p_s(z)$ and $p_r(z)$. We selected our
clusters for the CFHT sample based completely on the results of the
PDCS so to perform Monte Carlo simulations correctly would require
simulating the PDCS selection process.  Instead, we rely on the
simulations done by the authors of the PDCS catalog.  These
simulations show that, assuming the same elliptical galaxy type
K corrections as we use in \S 4.2, the PDCS has a 100\% probability
of successfully detecting a Richness Class 1 z = 0.4 in the $V$
band, with a 100\% probability of detection at z =0.55 for Richness
Class 2 or greater.

The analytical model for our probability of selection was
\begin{equation}
p_s(z_{est}|z) = \int_{z_{est} - \delta_z}^{z_{est} + \delta_z}
\frac{1}{\sqrt{2\pi}\sigma_z} \exp{\frac{-(z -
z_{est})^2}{2\sigma_z^2}} dz
\label{pzest}
\end{equation}
where $z_{est}$ is the estimated redshift from the PDCS catalog, $z$
is the actual redshift of a clusters, $\delta_z = 0.05$, or half the
interval between PDCS estimates redshifts.  The scatter in the
Gaussian, $\sigma_z = 0.063$, was determined in \S 3.1.4 (see Figure
\ref{zs_v_ph}).  We can invert the above equation to predict
$p(z|z_{est})$ if we make the assumption that all values of $z$ and
$z_{est}$ are equally likely.  We then multiply the probability above
with the simulation results from the PDCS.  The product of these two
probabilities determines $p(z)$ in equation (\ref{surveyvol}) for the
estimated redshift range in our sample.

The remaining quantities to be determined are the limits on the
integral in equation (\ref{surveyvol}) and the probability of
successfully measuring the redshift a cluster, $p_r(z)$.  We assume
because of the large number of spectra we collected per cluster
candidate ($\ge 40$) that $p_r(z)= 1.0$ for our $\Lambda_V \ge 40$ and
$0.3 \le z \le 0.5$ PDCS sample.  The main limit on the redshift range
is the sensitivity of our spectrograph and the blocking filters we
used.  The two blocking filters we used spanned from 0.20-0.56 and
0.43-0.61 in redshift.  Therefore, we chose $z_1 = 0.2$ and $z_2 =
0.61$.  Therefore, the resulting volume is $31.7^{+0.5}_{-0.8}\times
10^4$ h$^{-3}$ Mpc$^3$ for Richness Class 2 and greater clusters and
$21.9^{+0.5}_{-0.8}\times 10^4$ h$^{-3}$ Mpc$^3$ for Richness Class 1
clusters. 

The main uncertainty in the above estimate of the survey volume is the
actual shape of the $z_{estimated}-z$ probability distribution, which
we assumed to be Gaussian.  In \S 3.1.4 we test our assumption the
distribution is a Gaussian and that the mean of the distribution is
zero.  Nonetheless, as an alternative, we used the actual distribution
of redshift differences to compute a volume of $24.7\times 10^4$
h$^{-3}$ Mpc$^3$ for Richness Class 2 or greater clusters.  Therefore,
we conclude our error on the survey volume is, at most, $\sim$ 20\%.

\section{X-ray and Richness Space Density}

The combination of the ARC/KPNO sample and the CFHT sample will allow
us to measure the space density of PDCS clusters as a function of a
number of independent variables.  Here we will discuss the density
of PDCS clusters as a function of richness and X-ray luminosity,
leaving the mass for \S 6.

\subsection{Space Density as a Function of Richness}

The original PDCS catalog paper \citep{postman96} used the estimated
redshifts and richnesses to compute the space density of clusters as a
function of richness.  The authors of the PDCS found that the space
density for the PDCS was about five times that of the Abell catalog
for a given Richness Class and did not evolve with redshift.  The
original purpose behind the ARC/KPNO sample was to check this result.
In H99\nocite{holden99}, we found our survey to be in agreement with
\citet{postman96}.  Here we will improve on that work using our larger
sample of PDCS clusters.

To compute the space densities, we group the clusters by Richness
Class using the conversion between $\Lambda$ and Richness from
\citet{postman96} (see Table \ref{richdens} for the conversion).  We
used both the V and I band $\Lambda$ values separately.  We then
divided the number of clusters in each class by the volumes from \S
4.2 and \S 4.3.  The resulting number of clusters in each class and
densities are listed in Table \ref{richdens}, with the ARC/KPNO sample
and the CFHT sample listed separately.  

We plot the $\Lambda_V$ densities from Table \ref{richdens} in Figure
\ref{riches} in comparison with those from \citet{postman96}.  We find
a good match between the space density of clusters in our sample and
in the original PDCS.  This is not surprising given how good the
estimated redshifts are in the PDCS.  The only way we would have found
a serious discrepancy is if a large number of PDCS clusters turned out
to be false positive detections, which is not the case.

We compare our data with a sample of low redshift clusters of galaxies
created using the same algorithm, the sample of \citet{bramel99}.
This sample, also known as the Edinburgh-Durham Cluster Catalog 2
(EDCC2), used the algorithm of the PDCS to find clusters of galaxies
in the Edinburgh-Durham Galaxy Catalog.  We re-scaled the densities
from \citet{bramel99} to those appropriate for a $q_o = 0.1$ cosmology
and plot the resulting densities in Figure \ref{riches} as open
triangles.

In Figure \ref{riches}, we also plot the space density of clusters of
galaxies from the APM cluster catalog \citep{dalton97} using the
densities from \citet{croft99}.  The APM catalog is a low redshift ($z
\le 0.1$) catalog constructed using a related but somewhat different
technique then the PDCS.  We converted between the APM richness and
the $\Lambda$ richness using the values Table 1 and 2 of
\citet{dalton97} for the sample used in \citet{croft99}.

As demonstrated in Figure \ref{riches}, our space density measurements
for the PDCS are in good agreement with those from three different
analyses; EDCC2, APM and the original PDCS.  Given this strong
agreement between the high redshift measurements from the PDCS and the
results of the low redshift catalogs, we have confirmed the claim in
\citet{postman96} for no apparent evolution in the richness function
out to $z=0.5$.

\subsection{Space Density as a Function of X-ray Luminosity}

Using the volume estimates from \S 4, we can compute the space density
of clusters as a function of X-ray luminosity.  The main concern here
is that most of the PDCS clusters are not X-ray detections but are
upper limits.  Secondly, our sample is quite small compared with X-ray
selected cluster surveys.  Therefore, our X-ray luminosity function is
a check of the PDCS.  In other words, does it contain all the clusters
it should contain or does it miss clusters an X-ray selected sample
would fine?

We plot, in Figure \ref{nvl}, the cumulative number of detections
using solid circles.  We also plot, with open circles, the cumulative
distribution of upper limits and detections.  To construct this
cumulative distribution, we used a Kaplan-Meier product limit
estimator \citep{feigelson85, schmitt85}.  This is a parameter free
estimator that allows us to use both the upper limits and the
detections.  We use the formulation from \citet{schmitt85} which is
tailored for luminosity functions.
\begin{equation}
N\left(L_x > L_{x,i}\right) = 1 - \prod_{i,L_{x,i}>L_x}^{n}\ 
	\left(1 - \frac{d_i}{n_i}\right)
\label{kmest}
\end{equation}
where $L_x$ is the X-ray luminosity, $L_{x,i}$ is the X-ray luminosity
of the $i^{\rm th}$ cluster, $n_i$ is the number of clusters with
$L_{x,i} \le L_{x}$ and $d_i$ is the number of clusters with $L_{x,i}
= L_{x}$ so is either 1 or 0 for our sample.  As pointed out in
\citet{briel93}, this technique assumes that we know precisely the
values of the upper limits or of the detections.  Even our detections
are at a low ($\sim 3\sigma$) significance while there is a small
probability that a upper limit could be smaller than the actual
cluster luminosity.  Thus, this is a significant source of error.  

We also plot, using squares, the cumulative distribution function for
the ARC/KPNO sample.  Once again we plot both using only the
detections (solid squares) and both the detections and the upper
limits (open squares).  The luminosity function for the ARC/KPNO
survey is higher than that for the CFHT survey and higher than the
other published samples.  A likely explanation for this is we
under-estimated the volume surveyed for the ARC/KPNO sample.

For comparison, we plot in Figure \ref{nvl} the cumulative X-ray
luminosity functions from two low redshift samples: \citet{burns96}
(solid line) and the power law component of the Schechter luminosity
function from \citet{ebeling97} (dotted line).  We also plot power-law
fits \citet{burke97} (dot-dashed) and \citet{rosati98} ($+$ symbols),
two samples of the same redshift as our PDCS sample.  All the lines
are for $h=0.5$ and $q_o = 0.5$, so we recomputed our survey volume
and X-ray luminosities for these values.

There are two moderate to high redshift samples of optically selected
clusters of galaxies with X-ray luminosity functions.  One sample,
\citet{castander94}, has a lower space density at roughly the same
luminosities as our sample.  However, the sample of
\citet{castander94} is at much higher redshift than our sample, so the
disagreement with our luminosity function could be explained by
evolution.  The second sample is from \citet{bower94}.  In that
sample, almost all of the clusters of galaxies are detected, but the
luminosity function was estimated using both the detections and upper
limits.  That sample shows a decrease in the luminosity function as
well, though the mean redshift $\bar{z}=0.42$, quite close to our mean
redshift $\bar{z}=0.38$.  The more recent X-ray selected samples of
clusters of galaxies cast doubt on the results of \citet{bower94}, as
surveys such as \citet{burke97} and \citet{nichol99} find no evidence
for evolution in the same range of luminosities and redshifts.

Nonetheless, we confirm in Figure \ref{nvl} that the space density of
PDCS clusters is higher compared to X-ray selected samples of clusters
of galaxies.  Most of these clusters of galaxies, however, are not
detected as X-ray emitters.  This differs from other X-ray surveys of
optically selected clusters \citep{briel93, castander94,bower94,
burg94} which find that 50\% or more of the clusters in those samples
are detected.  When using only the detections in our sample, we find a
cumulative distribution that matches X-ray selected surveys for the
CFHT sample.  Therefore, it appears that the PDCS did not miss
clusters that an X-ray selected survey would find.

\section{Space Density as a Function of Mass}

We measured velocity dispersions of PDCS clusters of galaxies to
compare the resulting distributions with low redshift samples and as a
complement to other high redshift samples, such as the CNOC cluster
survey mentioned earlier.  

In Figure \ref{massf} we plot the cumulative distribution of cluster
masses in our sample.  We also plot the error bars on the mass
measurements.  The errors bars on the number density are comparable to
those found in our X-ray luminosity function in Figure \ref{nvl}.  We
also plot, on this diagram, the mass function of the CNOC cluster
survey using the velocity dispersions published in
\citet{carlberg96}.  We note here that we plot the entirety of our
sample, we do not try to create a subsample that is complete to some
limiting velocity dispersion as is done in \citet{carlberg97b} and
\citet{borgani99}.  However, we plot the X-ray luminosity and velocity
dispersion limited sample from \citet{carlberg97b}.  Though the two
cumulative distributions do not overlap, we see we fulfilled our goal
of creating a complementary survey to the CNOC survey. Most of
clusters are at a smaller mass than those found in the CNOC survey.
The combination of the two surveys now covers over three orders of
magnitude in space density.

The main problem with using our sample for measuring the mass function
of clusters of galaxies is the quality of our mass measurements.  As
we show above, the mass depends on the velocity dispersion cubed (see
equation \ref{msigma}), and most of our velocity dispersion
measurements have errors up to $\sim$50\%, yielding errors on the mass
of $\sim$150\%.  The question becomes how to take advantage of our
data, an almost complete subsample of the PDCS with velocity
dispersions, while minimizing the impact of our errors.

We elected to approach this problem by finding the best fitting
theoretical mass function.  We can then compare this best fit with the
results from other samples with much smaller errors.  In Figure
\ref{massf}, we plot three curves.  These are best fitting theoretical
mass functions to different data sets, generated using the model
discussed below.  The solid line is from the results of
\citet{bahcall97}, the dashed line is based on a fit to the mass
function of the CNOC survey by \citet{carlberg97b} and the dotted line
is our fit to the mass function of our data.  All three theoretical
curves are the result of letting two parameters $\Omega_m$, the mass
density of the universe, and $\sigma_8$, the normalization of the
power spectrum of mass fluctuations vary.  Each combination of these
two parameters predicts a different mass function, which is then
compared with the data to find a best fitting model.

Our theoretically expected mass functions are based on the
semi-analytical model originally from \citet{press74}.  We
specifically used the formalism outlined in \citet{viana96}.  Thus, we
used the following to model the density fluctuations spectrum
\begin{equation}
\sigma(R) = \sigma_8 \left( \frac{R}{8 h^{-1} {\rm Mpc}} \right)^{\gamma(R)}
\label{ps}
\end{equation}
where $\gamma(R)$ depends on $\Gamma$ and a few constants.  The
scale $R$ is related to the mass $M$ by the density $M = 4\pi R^3
\rho$. The form of $\gamma(R)$ is based on a fit to the spectrum of
fluctuations in cosmic microwave background and in the distribution of
galaxies.  The term $\sigma_8$ in equation (\ref{ps}) has a redshift
dependence that is different for different values of $\Omega_m$.  To
account for this dependence, we use the equations from
\citet{viana96}.

As we fixed $\Gamma = 0.23$ in equation (\ref{ps}), we have two free
parameters: $\Omega_m$ and $\sigma_8$, two of the parameters that we
said earlier are the most important for determining the number density
of virialized masses.  We could vary these independently, but instead
we chose to use the relation \( \sigma_8 = 0.60 \Omega_m^{-0.48 +
0.17\Omega_m} \) from \citet{girardi98b}.  This particular relation
was the result of a fit to a low redshift sample of clusters of
galaxies which makes comparing with the results of \citet{girardi98b}
easier.  As this relation results from a fit, instead of an analytical
derivation, there is an error associated with the fitted parameters on
the order of $\sim$10\%.  Other papers \citep{bahcall97,borgani99}
have found a similar relation with as the above but with constants
that vary on the order of $10\% - 20\%$ from the numbers above.

The Press-Schechter formalism predicts the space density of clusters
of galaxies as a function mass.  However, what we measured were the
velocity dispersions of clusters.  To convert the masses into velocity
dispersions, we used the estimate of \citet{borgani99}, or equation
(\ref{msigma}).  This allows us to fit our distribution of observed
velocity dispersions as a function of the above parameters using
the Press-Schechter formalism.

To estimate which value of $\Omega_m$ best agrees with our velocity
dispersion measurements, we used a maximum likelihood estimator.  In
particular, we used the approach of \citet{cash79} and
\citet{marshall83}, where we compute
\begin{equation}
S = -2 \ln L = -2 \sum_{i = 1}^{N} \ln\left[
\frac{dN(z_i,\sigma_{v,i}|\Omega_m,\sigma_8)}{dM} \frac{dV(\Lambda_i |
\Omega_m,\sigma_8)}{dz}   \right] - E(\Omega_m,\sigma_8)  
\label{maxlike}
\end{equation}
$z_i$ is the observed redshift a cluster, $\sigma_{v,i}$ is the observed
velocity dispersion of a cluster, $\Lambda_i$ is the observed
richness, $N$ is the total number of observed clusters in the sample,
and $E$ is the expected total number of clusters in the sample, a
double integral of the model over the redshift and mass range of the
sample.  We note here that for every different value of $\Omega_m$, we
recomputed the survey volume using the prescription in \S 4.3.

To compute $E$ in equation (\ref{maxlike}) requires that we have a
relation between the mass and the richness.  We used the richness-mass
relation in equation (\ref{richnessmass}).  We will assume that the
richness-mass relation does not evolve with redshift as we do not have
data in our sample to test for such evolution.  We note here, however,
that this is one of the most likely sources of systematic error in
addition to the large statistical error in the relation (see Figure
\ref{richvmass}).

To included the errors in our velocity dispersion measurements, we
convolved the $\frac{dV}{dz}\frac{dn}{dM}$ term in equation
(\ref{maxlike}) with a Gaussian of the width of the errors in the
velocity dispersion when computing the likelihood for a cluster given
a specific model.  The process is outlined in \citet{burke97} and
\citet{nichol97} and is similar to the process in \citet{borgani99}.

Using equation (\ref{maxlike}), we computed the likelihood for various
values of $\Omega_m$.  We found the best fit to be $\Omega_m = 0.32
\pm 0.08$ (68\% confidence limits for 1 parameter).  The small
apparent error on our estimate can be deceiving as it is a result of
the relation \(\sigma_8 = 0.60 \Omega_m^{-0.48 + 0.17\Omega_m} \).
Most of the clusters in our sample have masses at the value of the
mass enclosed in a 8 $h^{-1}$ Mpc sphere so our fit to the
mass function is most sensitive to $\sigma_8$, the normalization of
the power spectrum in equation \ref{ps}.  This also means that our
results are insensitive to changes in the parameters the change the
shape of of the mass function, such as $\Gamma$ and $\Omega_m$.
Therefore, when we force a specific relation between $\Omega_m$ and
$\sigma_8$, it appears we constrain $\Omega_m$ well, however the only
parameter we really can attempt to constrain is $\sigma_8$.  If
instead imposing the relation between $\Omega_m$ and $\sigma_8$, we
allow $\Omega_m$ and $\sigma_8$ to be independent, our 68\% confidence
limits for $\Omega_m$ span $0.1 \le \Omega_m \le 1.0$.

We plot in Figure \ref{massf} the cumulative mass function based on
the best fitting value of $\Omega_m$ to our data as the dotted line.
One important thing to note is that the mass function has a redshift
dependence.  We plot the theoretical mass functions for $z = 0.4$ but
our data and the CNOC data cover a redshift range of $0.2 \le z \le
0.6$.  Nonetheless, the dotted line, our best fit, agrees well with
our data and the CNOC cluster masses.

For comparison, the authors of \citet{carlberg97b} found a best fitting
$\Omega_m = 0.4$ and a best fitting $\sigma_8 = 0.75$ using the CNOC
cluster survey.  \citet{bahcall97}, which used the PDCS in part, found
a best fitting $\Omega_m = 0.2$ and $\sigma_8 = 0.85$.  These values
are only for comparison, each of them uses a different value of
$\Gamma$ as well as different relations between $\sigma_8$ and
$\Omega_m$.  In fact, in the re-analysis of \citet{carlberg97b},
\citet{borgani99} finds a range of allowed values of $\Omega_m$
depending on the relation between $\sigma_8$ and $\Omega_m$
used. However, \citet{borgani99} did find the best range to be $0.35
\le \Omega_m \le 1.0$.  The important result is that our results are
in agreement with the results of \citet{carlberg97b} and
\citet{borgani99} when we make similar specific assumptions for the
value of $\Gamma$ and for the relation between $\sigma_8$ and
$\Omega_m$.  

When we vary the relations we used, we greatly increase our
uncertainty.  For example, if we shift the scale of the richness-mass
relation (equation \ref{richnessmass}) by half a richness class, the
best fitting values range $0.26 \le \Omega_m \le 0.34$.  In the
simulations of the PDCS algorithm, it was found that some PDCS
clusters could be shifted as far as a whole Richness Class, which
would cause an even larger shift in the best fitting value of
$\Omega_m$.  Furthermore, the large scatter in the observed
richness-mass relation means we have a number of low mass clusters
contaminating our sample.

\section{Discussion and Conclusions}

In this paper we presented redshift measurements, velocity dispersions
and X-ray imaging data for a subset of the PDCS.  We aimed to construct
the first dataset of velocity dispersions of a subsample of the PDCS.
This dataset was constructed to be complementary to the CNOC cluster
survey and to be easily compared with lower redshift samples.  Our
goal was to use this dataset to determine the space
density of clusters of galaxies in the PDCS catalog as a function of
richness, X-ray luminosity, and mass.  We could then compare these
space densities with low redshift samples to test the apparent lack of
evolution in the PDCS cluster catalog and to compare the distribution
of masses with the expectations of cluster formation theory.

We began this process by collecting X-ray imaging data for a total of
thirty-one cluster candidates in the PDCS.  We followed up the X-ray
imaging data with spectroscopy of PDCS cluster candidate galaxies.  We
successfully measured redshifts for seventeen PDCS clusters (fifteen
of which we have X-ray imaging data for).  The PDCS clusters of
galaxies for which we have redshifts are part of two samples with
different selection criteria.  One sample, the ARC/KPNO sample, was
selected on the number of $V-I \ge 1$, $V \le 21$ galaxies in a
2\farcm 5 circle centered on the PDCS cluster candidate position.
This provided a sample selected independently of many of the derived
cluster parameters in the PDCS (such as richness, estimated redshift,
{\em etc}.)  Our second sample, the CFHT sample, was selected using
cluster parameters from the catalog, specifically all clusters in the
X-ray imaging data with $\Lambda_V \ge 40$ and $0.3 \le z_{estimated}
\le 0.5$ where $\Lambda_V$ is the richness estimate of the PDCS in the
V filter.  This second sample was constructed to mimic the selection
of the ESO Nearby Abell Cluster Survey in that we selected clusters of
Richness Class 1 or greater \citep{katgert96}.  

Given these two samples of redshifts, we then derived
redshifts for the clusters, and, when possible, velocity dispersions
based on the galaxy redshifts.  We fit a Gaussian plus flat background
using a maximum likelihood estimator to measure the velocity
dispersions.  This approach has the advantage of using the galaxy
velocities directly and is completely automated.  However, as we made
an explicit choice on the shape of the velocity distribution, we are
biased towards over-estimating the velocity dispersion of clusters
with complicated dynamics.  We tested our approach using the published
CNOC velocities and found that for seven out of the eight clusters we
measured the same velocity dispersions and the same size errors,
$\sim$ 10\%.  The one exception was MS 1512+36, a dynamically
complicated cluster with two secondary peaks in the velocity
distribution.  Most of the clusters in our sample had a much smaller
number of cluster members than the CNOC sample.  Therefore, our
velocity dispersions had much larger errors, up to $\sim$ 50\%.

Using the redshifts, we measured X-ray luminosities or upper limits
for thirty-one PDCS cluster candidates.  We chose to measure our X-ray
luminosities within an aperture that contains a fixed fraction of the
total flux, using a $\beta$ model to calculate the size of that
aperture.  For those sixteen PDCS cluster candidates for which we did
not have spectroscopic redshifts, we used the estimated redshifts of
the PDCS to compute the apertures.  Twenty-four of the candidates in our
sample were not X-ray detections.  For those, we assigned
$3\sigma$ upper limits on the flux and luminosity.

\begin{figure}
\begin{center}
\includegraphics[height=5in]{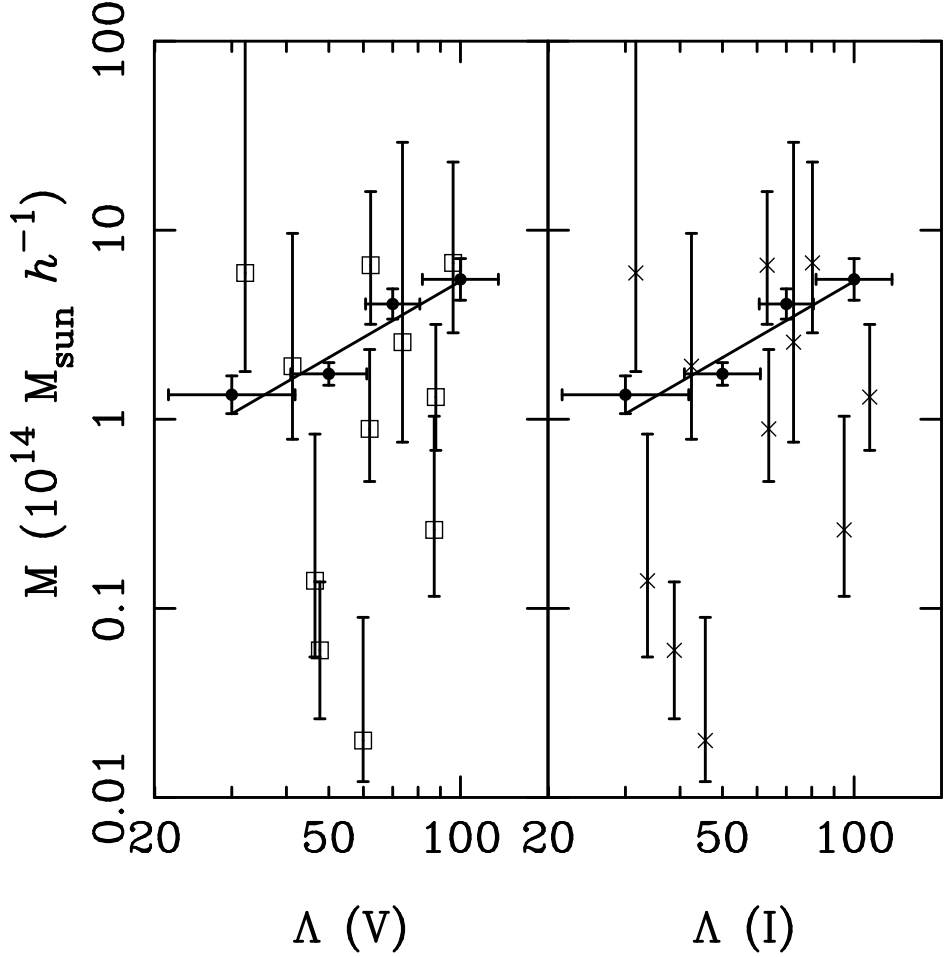}
\end{center}
\caption[holden.fig5.ps]{Mass as a function of richness.  We use
the mass from equation (\ref{msigma}) and the PDCS $\Lambda$ richness.
The V band richness is plotted with open squares in the left panel
while the I band richness is plotted with crosses in the right panel.
Also, we plot the mass-richness relation from Girardi \etal (1998a).
That relation is plotted using solid circles.  The vertical error bars
represent the scatter around the mean of the sample at that Richness
Class.  The x-axis error bars represent the range in PDCS richness for
the Richness Class.  The Richness Classes in this plot are (from left
to right) 0, 1, 2, and $\ge$ 3.  We plot, as a line, our fit to the
Girardi \etal (1998a) richness-mass relation.
\label{richvmass}}
\end{figure}

\begin{figure}
\begin{center}
\includegraphics[height=5in]{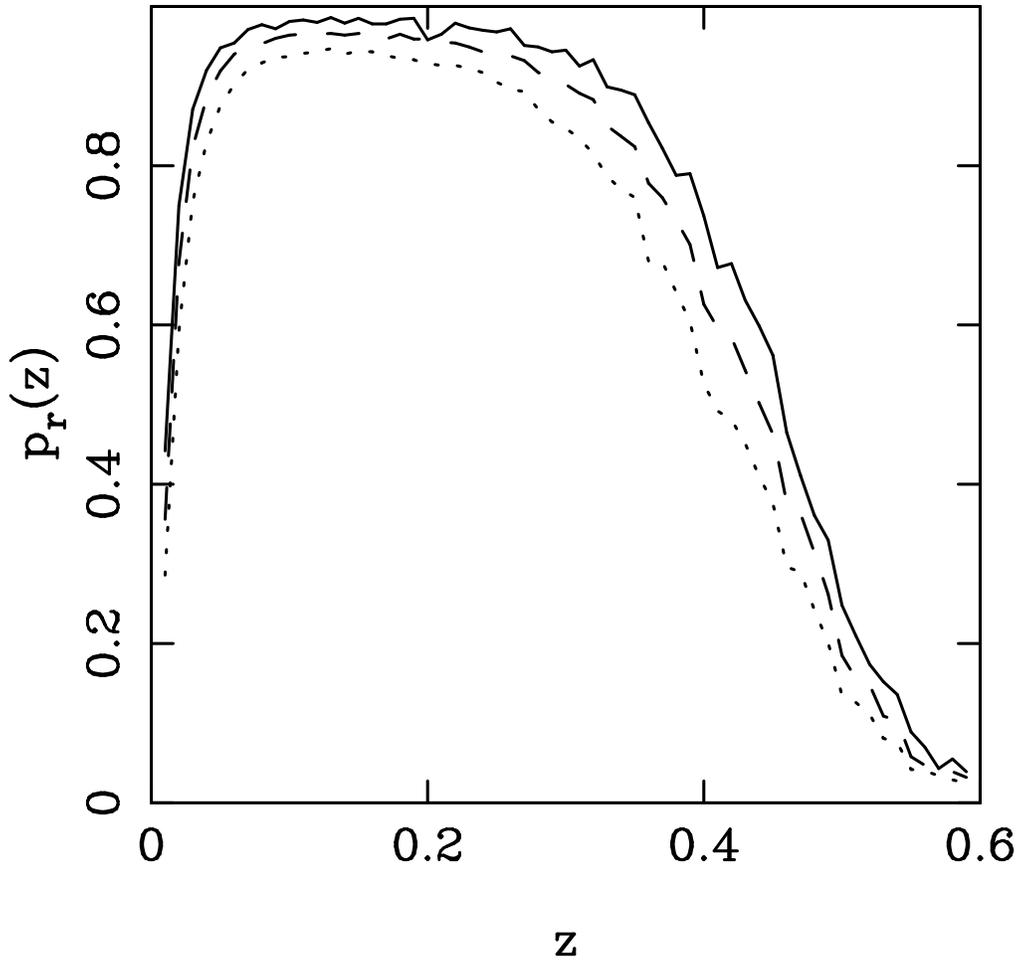}
\end{center}
\caption[holden.fig6.ps]{Probability of successfully
measuring a redshift for the ARC/KPNO sample as a function of
redshift.  The three lies are for Richness Class 3
(solid), Richness Class 2 (dashed) and Richness Class 1 (dotted).
\label{probrvz}}
\end{figure}

\begin{figure}
\begin{center}
\includegraphics[height=5in]{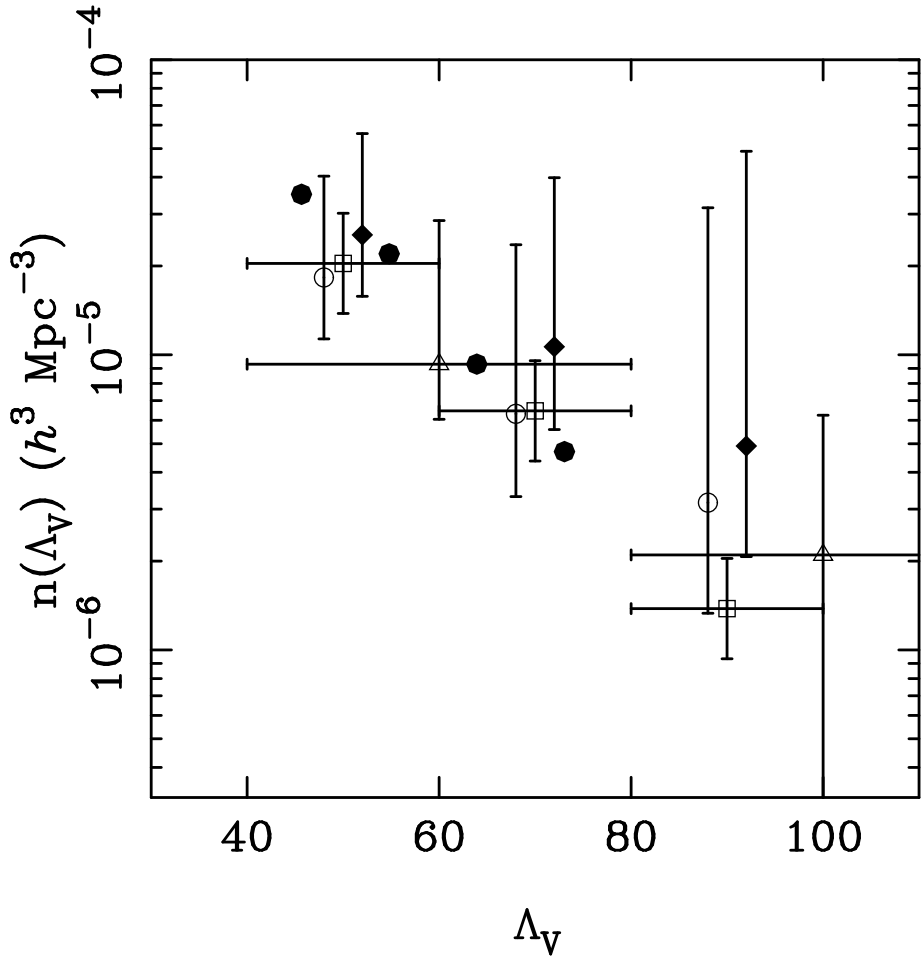}
\end{center}
\caption[holden.fig7.ps]{ Space density as a function of Richness
for the PDCS clusters in the original catalog (open squares), from the
ARC/KPNO sample (solid diamonds), and from the CFHT sample (open
circles).  The vertical error bars are the 68\% confidence limits for
a Poisson distribution from Gehrels (1986).  The open triangles
represent the results from Table 2 of Bramel \etal (2000).  We
also plot with solid circles the space density from the APM cluster
catalog.  The error bars for the original catalog are as given in
Postman \etal (1996).  The horizontal error bars give the range in
Richness covered by the different Richness Classes.  We note here that
Bramel \etal (2000) use a different relation between $\Lambda$ and
Abell Richness Class.
\label{riches}}
\end{figure}

When we compare our masses derived from the velocity dispersions to
the richnesses in the PDCS catalog in Figure \ref{richvmass}, we find
that \slantfrac{1}{3} of our masses are lower than the expected values
based on the richnesses.  For example, we have in our sample two
Richness Class 3 clusters with velocity dispersions of only $\sim$500
km s$^{-1}$, with implied masses around $1-3 \times 10^{14}$ $h^{-1}$
M$_{\sun}$.  We find that the relation between masses and richnesses
in the PDCS sample differs from the Abell mass-richness relation from
\cite{girardi98b} at the 99.5\% confidence limit.  This observed
difference comes from these low mass systems with high richnesses.

In \S 4.2 and 4.3 we derived the volumes surveyed for our two samples.
We find that these volumes are accurate to around $\sim$ 20\% except
for changes in $q_o$.  These are much smaller errors than our errors
based on our sample size.  We used these volumes to derive the space
density of PDCS clusters as a function of richness, X-ray luminosity
and mass.  If we examine the richness function of the clusters in our
sample, we find space densities in line with those of the whole of the
PDCS.  As we mentioned before, this result is not surprising given how
accurate the estimated redshifts are for the clusters in our sample
and that most of the candidates in the PDCS are true over-densities of
galaxies.  Based on this result, we find that the PDCS agrees with the
space densities for lower redshift catalogs of clusters of galaxies
like the EDCC2 of \citet{bramel99} and the APM results from
\citet{croft99}.  However, this raises the question of why all of
these catalogs find a much higher space density of clusters of
galaxies than the Abell catalog at ostensibly the same richnesses?

\begin{figure}
\begin{center}
\includegraphics[height=5in]{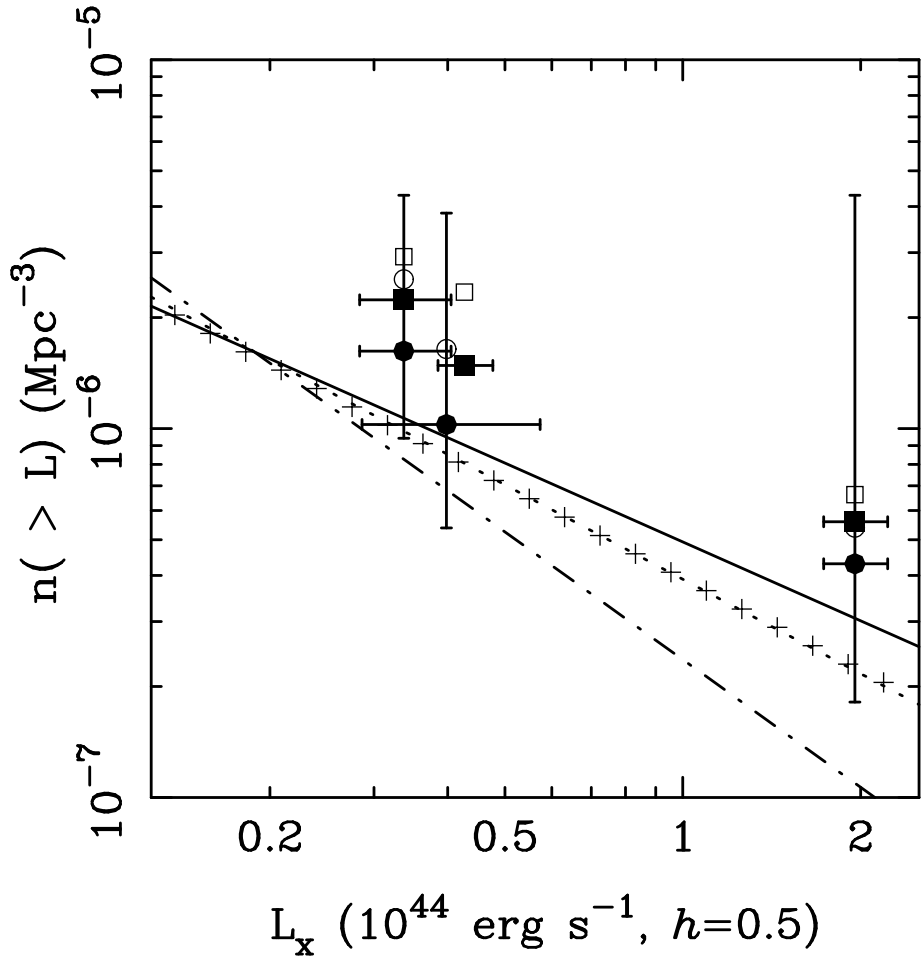}
\end{center}
\caption[holden.fig8.ps]{
Cumulative X-ray luminosity function of PDCS clusters in the CFHT
sample using both the detections (solid circles) and upper limits
(open circles).  Our results from the ARC/KPNO sample are plotted
using squares with solid representing the detections and open
representing the combination of detections and upper limits.  We plot
the parametric distributions from {Burns} {\em et~al.}
(1996) (solid line), {Ebeling} {\em et~al.}
(1997) (dotted line), {Burke} {\em et~al.}
(1997) (dot-dashed line) \& {Rosati} {\em et~al.}
(1998) ($+$ symbols).  The error bars are for a
Poisson distribution at each point in the cumulative distribution of
detections.  The luminosity errors are 68\% confidence limits from
Table \ref{xrayresults}.\label{nvl}}
\end{figure}

To try to answer this question, we first look at a different
distribution function, the X-ray luminosity function.  What is
interesting is that the X-ray luminosity function of PDCS clusters of
galaxies is quite in line with those based on X-ray selected samples,
such as \citet{burke97} and \citet{rosati98} as well as those based on
other optically selected samples such as \citet{burns96}.  This means
that it is unlikely that the PDCS missed any X-ray emitting clusters
of galaxies.  However, considering that most of the clusters of
galaxies are not detected in our X-ray data, those clusters must have
quite faint luminosities ($ \le 10^{43}$ erg s$^{-1}$, see Figure
\ref{nvl}).  The clusters at those luminosities in the
optically-selected, low redshift sample of \cite{burns96} are quite
poor, being either Richness Class 0 Abell clusters or groups from the
catalog of \citet{white99}.

If we examine the mass function, once again we find that the
distribution of masses in our sample matches that of other samples.
We chose to compare our mass function with those of other samples by
finding which cosmological parameters are most likely to produce the
distribution of masses in our sample.  We find the best fitting
$\Omega_m$ to be quite in line with those from the CNOC cluster survey
if we assume both the $\sigma_8 - \Omega_m$ relation and the
richness-mass relation from \citet{girardi98b}.
By the theoretical mass function that best fits our data, we see that
most of our clusters should have masses on the order of $\sim 10^{14}$
h$^{-1}$ M$_{\odot}$, which is in agreement with our data, and is the
mass range found for Richness Class 0 and 1 clusters in Figure
\ref{richvmass}.

Based on our data, we find that the PDCS finds clusters of galaxies
have distributions of mass and X-ray luminosity that match those of
other samples.  This leaves the richness function as the discrepant
function, but only when compared with the Abell catalog.  Other
optically selected catalogs of clusters of galaxies find a much higher
space density of objects than the Abell catalog, such as the APM
cluster catalog \citep{dalton97} or the Edinburgh-Durham Cluster
Catalog 2 \citep{bramel99}.  Therefore, it is likely that either the
Abell catalog is strongly incomplete by a factor of three or four, a
rather unlikely occurrence, or that there is a mismatch between the
various richness measures used by the automated catalogs and the Abell
Richness.  These ideas will be tested with future cluster catalogs
that cover a large area of the sky, such as the DPOSS \citep{gal99}
and SDSS cluster catalogs.

\begin{figure}
\begin{center}
\includegraphics[height=5in]{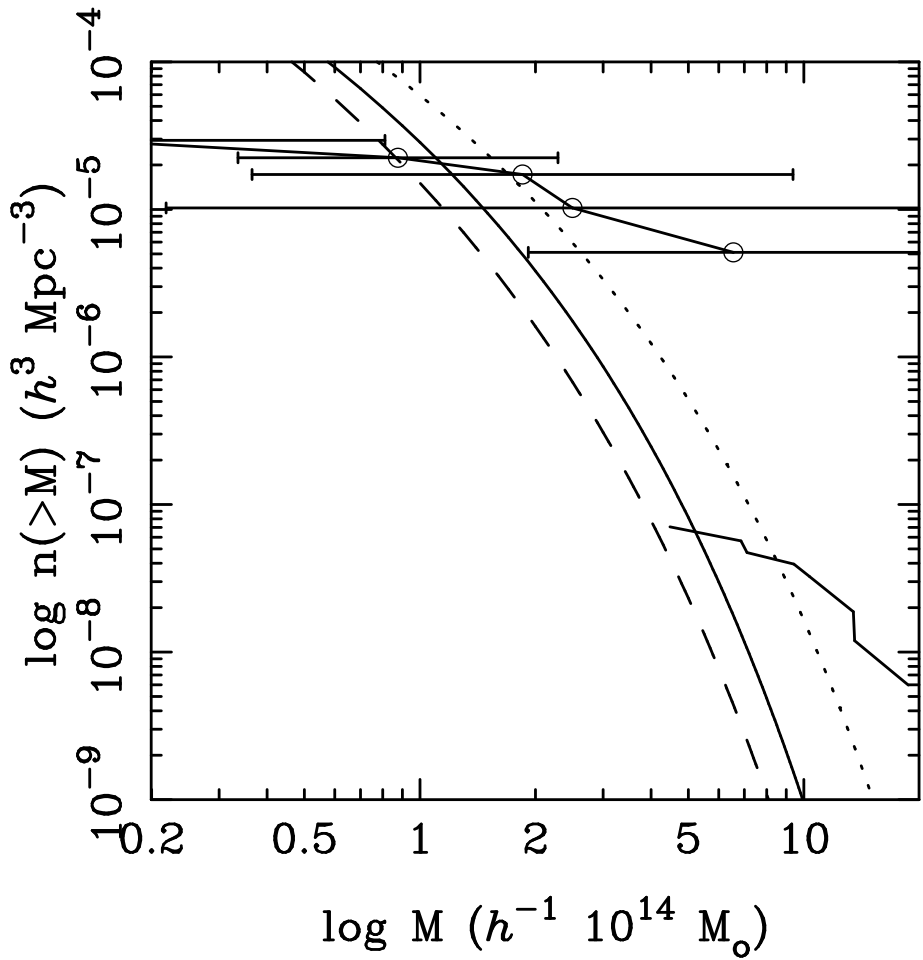}
\end{center}
\caption[holden.fig9.ps]{ Cumulative space density of PDCS clusters
as a function of mass from the CFHT sample.  The points with error
bars represents our cumulative distribution function.  The solid curve
the lower right is the cumulative mass distribution for the CNOC
cluster survey.  The smooth solid line represents the best fit from
{Bahcall}, {Fan}, \& {Cen} (1997), the dashed line
is from Carlberg \etal (1997), a fit to the CNOC
cluster sample, and the dotted line is the $\Omega_m = 0.32$ and
$\sigma_8 = 0.97$ line from the text.  All of the theoretical mass
functions are calculated for a redshift of 0.4, the mean of the range
of allowed redshifts.  The volumes for the CNOC and PDCS mass
functions are computed for $q_o = 0.16$.
\label{massf}}
\end{figure}

The promise of optically selected catalogs has always been that, given
the relative ease of collecting large amounts of imaging data,
catalogs with literally thousands of clusters of galaxies could be
constructed.  Our investigations of the PDCS shows that modern
sophisticated cluster finding algorithms do an excellent job of
finding over-densities of galaxies and correctly estimating their
apparent redshift.  From Figure \ref{massf}, we can see that it will
be possible to use optically selected clusters of galaxies to study
cluster formation and evolution.  The challenge will lie in improving
the contrast for truly massive systems, to decrease the contamination
we see in Figure \ref{richvmass}.  Machine based richnesses appear to
be only an incremental improvement over counts by eye \citep[for a
comparison of richness to mass for the APM and Abell
catalogs]{alonso99}.  A possible solution could be in using such
techniques as a color-based or photometric redshift selection, see for
example \citet{kepner99} or \citet{gal99}.  The approaches have the
promise of reducing the number of contaminating field galaxies in
cluster candidates.  A second solution would be to replace a simple
count of the number of galaxies with new measures of the total galaxy
content such as the total luminosity from \citet{adami98},
\citet{fritsch99} or \citet{miller00}.  Catalogs using the combination
of these two techniques should have the observed optical galaxy signal
much more strongly correlated with mass.  With such improvements and
the new large area surveys such as the SDSS and DPOSS, optically
selected clusters can be used to probe the mass function over 4 or 5
orders of magnitude in density and, thus, be used to explore the
formation of the most massive gravitationally bound objects observed.
 
This paper represents the submitted version of BH's dissertation.  I
would like to thank a number of people for reading over various
versions of this work.  This list includes Rich Kron, who read the
whole thing at least twice, and my collaborators: Christophe Adami,
Francisco Javier Castander, Lori Lubin, Robert Nichol, and
A. Katherine Romer.  I would also like to thank the anonymous referee
for improving this paper.  For help in understanding statistics, I
thank Erik Reese for many useful pointers and Carlo Graziani who spent
hours explaining Maximum Likelihood parameter estimation and the
theory behind confidence limits.  I would like to credit Gil Holder
for correcting many of my misunderstandings of the Press-Schechter
formulation and for testing my predictions of the cluster mass
function.  Daniel Reichart also helped with my understanding of
Press-Schechter theory as well as all of the pitfalls of converting
observational quantities like richness into the masses predicted from
theory.  Finally, I would like to thank Chris Metzler and Martin White
for long explanations during the Cluster Eating Group about cluster
formation and the spherical collapse model, I would like to thank
Chris especially for his long patient explanations of relating
theoretical predictions to observations.

For scientific guidance, I would like to thank Rich Kron for teaching
so well his particular way of looking at observational data.  His
intuition and ability to quickly get to the heart of a complicated
collection of numbers served me well, despite my writing a thesis in a
field that is not his specialty.  I owe a great debt to Bob Nichol for
teaching me, on a daily basis, how to start, progress, and complete a
scientific project.  Without his ideas and guidance, I would have been
a graduate student much longer.  I only hope I can learn how to form
the ideas that he does.  Kath Romer has taught me almost all of the
X-ray astronomy I know, and good deal of the optical astronomy as
well.  She also patiently waded through the horrid language of my
first paper, molding it into something that could be read and
understood.  Francisco Castander had many long conversations with me
about cluster X-ray emission, stellar population synthesis, and all
sorts of other fields of astronomy.  I would like to thank Lori Lubin
taking the time to repeatedly dig through all of her old notes to answer
my obscure questions about the PDCS.  I would also like to thank her
for providing very insightful comments on earlier drafts of this
dissertation and my other papers.  In addition, I would like to thank
Mel Ulmer for scientific and financial support all throughout my
graduate career, as well as a fresh perspective on many different
scientific problems.  

This project was funded in part by NASA grant NAG5-3202.  BH was
partially supported by the Center for Astrophysical Research in
Antarctica, a National Science Foundation Science and Technology
Center, by NASA GO-06838.01-95A and by NSF AST-9256606.  This research
was supported through NASA ADP grant NAG5-2432 (at Northwestern
University) and NASA LTSA grant NAG5-6548 (at Carnegie Mellon
University).

\pagestyle{empty}

\tablenum{1}
\begin{deluxetable}{lllllrrrr}
\tablecaption{PDCS Cluster Candidates X-ray Data: Observed
Quantities.\label{rawxraydata}}
\tablewidth{480pt}
\tablehead{\colhead{PDCS \#\tablenotemark{a}} & \colhead{$\alpha$} & \colhead{$\delta$} 
& \colhead{z\tablenotemark{b}} & \colhead{Aper.\tablenotemark{c}} & \colhead{Time} & \colhead{Cts} 
& \colhead{Background}& \colhead{Fraction\tablenotemark{d}} \\ 
\colhead{} & \colhead{(J2000)} & \colhead{(J2000)} & \colhead{}
&\colhead{} & \colhead{(s)} & \colhead{} &  \colhead{Cts}  
&\colhead{} \\ 
}
\startdata
01 & 00 29 03.0 & +05 01 24 & 0.600 & 2\farcm 56 &  4224.0 &  35 & 20.7 & 0.80 \\
02 & 00 29 28.4 & +05 04 03 & 0.246 & 4\farcm 17 &  4897.4 &  82 & 76.5 & 0.80 \\
03 & 00 28 36.6 & +05 07 44 & 0.600 & 2\farcm 42 &  5465.8 &  31 & 29.7 & 0.80 \\
04 & 00 29 11.1 & +05 08 55 & 0.549 & 2\farcm 76 &  4660.1 &  11 & 11.2 & 0.53 \\
05 & 00 27 38.6 & +05 09 46 & 0.211 & 4\farcm 28 &  6328.5 &  86 & 84.0 & 0.79 \\
06 & 00 29 52.2 & +05 12 36 & 0.400 & 3\farcm 98 &  4333.9 &  75 & 59.7 & 0.80 \\
08 & 00 28 31.4 & +05 18 01 & 0.600 & 2\farcm 50 &  3666.4 &  22 & 25.9 & 0.76 \\
29 & 09 53 12.1 & +47 08 58 & 0.400 & 3\farcm 18 & 17183.2 & 194 & 196.6 & 0.80 \\
30 & 09 54 46.3 & +47 10 48 & 0.259 & 4\farcm 19 & 14210.8 & 290 & 269.6 & 0.78 \\
31 & 09 53 39.5 & +47 12 58 & 1.100 & 2\farcm 26 & 17687.0 & 120 & 90.6 & 0.80 \\
32 & 09 52 29.0 & +47 17 49 & 0.211 & 4\farcm 33 & 18326.3 & 328 & 300.0 & 0.77 \\
33* & 09 52 13.1 & +47 16 48 & 0.553 & 2\farcm 57 & 17446.5 & 107 & 77.2 & 0.69 \\
34 & 09 55 09.1 & +47 29 55 & 0.333 & 3\farcm 39 & 12202.6 & 211 & 193.4 & 0.80 \\
35 & 09 52 31.2 & +47 36 27 & 0.600 & 2\farcm 36 & 17409.9 & 125 & 102.0 & 0.80 \\
36* & 09 53 53.7 & +47 40 15 & 0.251 & 3\farcm 80 & 14564.7 & 369 & 261.6 & 0.80 \\
37 & 09 51 41.5 & +47 41 30 & 0.600 & 2\farcm 43 & 14651.7 & 117 & 90.8 & 0.80 \\
38 & 09 51 09.9 & +47 43 54 & 0.332 & 3\farcm 34 & 14432.2 & 154 & 159.4 & 0.80 \\
39 & 09 51 25.2 & +47 49 50 & 0.469 & 2\farcm 91 & 11528.6 & 113 & 127.0 & 0.80 \\
40* & 09 53 25.6 & +47 58 55 & 0.203 & 4\farcm 63 & 10704.2 & 331 & 161.3 & 0.76 \\
41* & 09 54 16.9 & +47 58 41 & 0.700 & 2\farcm 98 &  9667.9 & 102 & 71.6 & 0.77 \\
42 & 09 53 54.2 & +48 00 04 & 0.900 & 2\farcm 82 &  9401.2 &  60 & 51.3 & 0.72 \\
43 & 09 52 15.1 & +47 57 44 & 0.200 & 4\farcm 65 & 11699.9 & 279 & 259.1 & 0.78 \\
44 & 09 52 18.6 & +48 02 32 & 1.100 & 2\farcm 92 &  9925.8 &  76 & 91.2 & 0.78 \\
45* & 09 54 38.8 & +47 15 59 & 0.400 & 3\farcm 18 & 15273.1 & 228 & 161.4 & 0.79 \\
\enddata

\end{deluxetable}

\tablenum{1}
\begin{deluxetable}{lllllrrrr}
\tablecaption{PDCS Cluster Candidates X-ray Data: Observed
Quantities.}
\tablewidth{480pt}
\tablehead{\colhead{PDCS \#\tablenotemark{a}} & \colhead{$\alpha$} & \colhead{$\delta$} 
& \colhead{z\tablenotemark{b}} & \colhead{Aper.\tablenotemark{c}} & \colhead{Time} & \colhead{Cts} 
& \colhead{Background}& \colhead{Fraction\tablenotemark{d}} \\ 
\colhead{} & \colhead{(J2000)} & \colhead{(J2000)} & \colhead{}
&\colhead{} & \colhead{(s)} & \colhead{} &  \colhead{Cts}  
&\colhead{} \\ 
}
\startdata
57 & 13 23 47.3 & +30 03 31 & 0.459 & 3\farcm 85 & 15357.7 & 363 & 331.2 & 0.80 \\
59 & 13 24 48.8 & +30 11 36 & 0.751 & 2\farcm 51 & 16828.8 & 160 & 152.0 & 0.80 \\
60 & 13 23 39.0 & +30 12 12 & 0.200 & 4\farcm 71 & 14693.9 & 408 & 502.7 & 0.80 \\
61 & 13 27 07.4 & +30 18 01 & 0.300 & 4\farcm 41 & 13924.4 & 252 & 243.3 & 0.72 \\
62* & 13 23 39.0 & +30 22 26 & 0.462 & 2\farcm 82 & 15467.2 & 345 & 196.2 & 0.80 \\
63* & 13 24 20.6 & +30 12 52 & 0.697 & 2\farcm 53 & 16252.9 & 209 & 121.2 & 0.80 \\
64 &  13 26 22.3 & +30 15 20 & 1.000 & 2\farcm 79 & 19129.5 & 177 & 151.3 & 0.80 \\
\enddata

\tablenotetext{a}{The PDCS identification number, clusters marked with
an * are X-ray detections.}
\tablenotetext{b}{Redshift used to compute the aperture size}
\tablenotetext{c}{Aperture that contains 80\% of the total flux for the
model specified in the text}
\tablenotetext{d}{The fraction of the model total flux expected to fall
with in the aperture.}

\end{deluxetable}

\tablenum{2}
\begin{deluxetable}{lllccccr}
\tablecaption{PDCS Candidates in CFHT Sample.\label{cfhtobs}}
\tablewidth{0pt}
\tablehead{\colhead{PDCS \#}  &\colhead{$\alpha$} &\colhead{$\delta$} &
\colhead{$z_{estimated}$\tablenotemark{a}} &
\colhead{$\Lambda_V$\tablenotemark{a}} & \colhead{\# of
Masks\tablenotemark{b}} & \colhead{z range\tablenotemark{c}} \\ 
\colhead{} & \colhead{(J2000.0)} & \colhead{(J2000.0)} & \colhead{}
& \colhead{} & \colhead{} & \colhead{} \\ 
}
\startdata
04 & 00 29 11.1 & +05 08 55 & 0.5 & 87.1 & 2 & 0.43-0.61 \\ 
16 & 02 28 26.5 & +00 32 20 & 0.5 & 87.8 & 2 & 0.43-0.61 \\
30 & 09 54 46.3 & +47 10 48 & 0.3 & 46.5 & 2 & 0.20-0.56 \\
32 & 09 52 29.0 & +47 17 49 & 0.3 & 32.2 & in 33 & 0.43-0.61 \\
33 & 09 52 13.1 & +47 16 48 & 0.5 & 41.3 & 2 & 0.43-0.61 \\
34 & 09 55 06.1 & +47 29 56 & 0.3 & 62.0 & 2 & 0.20-0.56 \\
38 & 09 51 09.9 & +47 43 55 & 0.3 & 47.7 & 2 & 0.20-0.56 \\
39 & 09 51 25.2 & +47 49 50 & 0.6 & 62.3 & in 38 & 0.20-0.56 \\
45 & 09 53 48.4 & +47 57 57 & 0.4 & 36.6 & in 30 & 0.20-0.56 \\
57 & 13 23 47.9 & +30 03 24 & 0.5 & 73.7 & 2 & 0.43-0.61 \\
61 & 13 27 02.4 & +30 18 14 & 0.3 & 33.2 & 1 & 0.20-0.56 \\
62 & 13 23 42.2 & +30 22 38 & 0.4 & 96.2 & 2 & 0.43-0.61 \\
67 & 16 93 49.4 & +41 11 13 & 0.5 & 59.9 & 1 & 0.43-0.61 \\
\enddata

\tablenotetext{a}{  Column four lists the estimated redshift and
column five lists the richness which were used to select clusters for
the observation.}
\tablenotetext{b}{The number masks used for measuring redshift of
cluster members.  If a cluster is noted as ``in \#'', that cluster was
observed using the masks for another PDCS cluster.  The number
specifies which cluster that is.}
\tablenotetext{c}{The redshift ranges of the blocking
filter used, see {Adami} {\em et~al.} 2000 for how these ranges were
determined.} 
\end{deluxetable}

\tablenum{3}
\begin{deluxetable}{lcccc}
\tablecaption{PDCS cluster redshifts for clusters only in ARC/KPNO
sample\label{otherzs}}
\tablewidth{0pt}
\tablehead{\colhead{PDCS \#} & \colhead{z\tablenotemark{a}} &
\colhead{\# of galaxies\tablenotemark{b}} & \colhead{$\Lambda_V$} &
\colhead{z$_{estimated}$} \\   
}
\startdata
02 & 0.2463 & 3/7 & 44.4 & 0.4 \\
05 & 0.2112 & 3/3 & 31.8 & 0.2 \\
12 & 0.2634 & 6/8 & 75.5 & 0.3 \\
23 & 0.1288 & 6/7 & 43.6 & 0.2 \\
36 & 0.2505 & 4/6 & 54.7 & 0.3 \\
40 & 0.2028 & 5/5 & 28.3 & 0.2 \\
\enddata
\tablenotetext{a}{``Best'' redshift from {Holden} {\em et~al.} 1999.}
\tablenotetext{b}{Fraction of total redshifts observed within 1500 km
s$^{-1}$ of the ``best'' redshift.}
\end{deluxetable}

\tablenum{4}
\begin{deluxetable}{lllrllr}
\tablecaption{Redshifts and Velocity Dispersions of CFHT
sample.\label{zmeans}}
\tablewidth{0pt}
\tablehead{\colhead{PDCS \#} & \colhead{${\rm z_{biweight}\tablenotemark{a} }$} &
\colhead{$\sigma_{v,\rm biweight, obs}$\tablenotemark{a}} &
\colhead{$\sigma_{v,\rm biweight, rest}$\tablenotemark{a,b}} & \colhead{${\rm
z_{ML} }$\tablenotemark{c}} & \colhead{$\sigma_{v, \rm
ML,obs}$\tablenotemark{c}} & \colhead{$\sigma_{v, \rm
ML,rest}$\tablenotemark{b,c}} \\ 
\colhead{} & \colhead{} & \colhead{} & \colhead{(km s$^{-1}$)}  &
\colhead{} & \colhead{} & \colhead{(km s$^{-1}$)} \\ 
}
\startdata
04 & 0.5486 &  0.002288 &  443 & 0.5486 & 0.001900 &  368$^{+169}_{-97}$\\
16 & 0.3984 &  0.003149 &  675 & 0.3984 & 0.002820 &  604$^{+178}_{-130}$\\
30 & 0.3311 &  0.001957 &  441 & 0.3314 & 0.000515 &  280$^{+167}_{-88}$\\
32 & 0.2106 &  0.004507 & 1111 & 0.2105 & 0.003824 &  947$^{+167}_{-88}$\\
33 & 0.5530 &  0.004414 &  852 & 0.5530 & 0.003689 &  712$^{+385}_{-212}$\\
34 & \nodata&  \nodata  & \nodata& 0.3327 & 0.002322 &  522$^{+167}_{-112}$\\ 
38 & 0.3316 &  0.001169 &  263 & 0.3322 & 0.000444 &  101$^{+28}_{-24}$\\
39 & 0.4691 &  0.006063 & 1237 & 0.4692 & 0.005149 & 1051$^{+314}_{-252}$\\
57 & 0.4586 &  0.004274 &  878 & 0.4588 & 0.003730 &  767$^{+628}_{-271}$\\
62 & 0.4619 &  0.005858 & 1201 & 0.4621 & 0.005162 & 1059$^{+434}_{-300}$\\
67 & 0.4675 &  0.000896 &  183 & 0.4675 & 0.000760 &  155$^{+68}_{-31}$\\
\enddata

\tablenotetext{a}{ The values noted by biweight
are derived from the first pass through the data.}
\tablenotetext{b}{  For both the maximum likelihood fits
and the biweight estimators, we give the rest-frame velocity
dispersion in km s$^{-1}$.}
\tablenotetext{c}{  The values noted by
ML are from the maximum likelihood fits of a Gaussian plus flat
background galaxy distribution.}
\end{deluxetable}

\tablenum{5}
\begin{deluxetable}{lll}
\tablecaption{PDCS Cluster Candidate X-ray Data: Physical
Quantities\label{xrayresults}}
\tablewidth{0pt}
\tablehead{
\colhead{PDCS \#\tablenotemark{a}} & \colhead{Flux\tablenotemark{b}} &
\colhead{Luminosity\tablenotemark{b}} \\ 
\colhead{} & \colhead{($10^{-14}$ erg s$^{-1}$ cm$^{-2}$)} &
\colhead{($10^{43}$ erg s$^{-1}$)} \\ 
}
\startdata
01  & 5.4 & 2.4 \\ 
02  & 9.7 & 0.7 \\ 
03  & 5.4 & 2.5 \\ 
04  & 6.0 & 2.3 \\ 
05  & 7.7 & 0.4 \\ 
06  & 9.3 & 1.8 \\ 
08  & 6.0 & 2.7 \\ 
29  & 4.1 & 0.8 \\ 
30  & 5.5 & 0.4 \\ 
31  & 2.6 & 4.4 \\ 
32  & 4.8 & 0.2 \\ 
33* &  3.1$^{+1.1}_{-1.0}$ & $ 1.2^{+0.4}_{-0.4}$  \\ 
34  & 5.3 & 0.7 \\ 
35  & 2.9 & 1.3 \\ 
36* &  12.5$^{+2.3}_{-2.2}$ & $ 0.9^{+0.2}_{-0.2}$  \\ 
37  & 3.1 & 1.4 \\ 
38  & 4.3 & 0.6 \\ 
39  & 4.4 & 1.2 \\ 
40* &  24.1$^{+2.7}_{-2.5}$ & $ 1.1^{+0.1}_{-0.1}$  \\ 
41* &  4.3$^{+1.5}_{-1.3}$ & $ 2.7^{+0.9}_{-0.8}$  \\ 
42  & 3.7 & 4.0 \\ 
43  & 7.0 & 0.3 \\ 
44  & 5.0 & 8.3 \\ 
45* &  6.2$^{+1.5}_{-1.4}$ & $ 1.2^{+0.3}_{-0.3}$  \\ 
\enddata
\end{deluxetable}

\tablenum{5}
\begin{deluxetable}{lll}
\tablecaption{PDCS Cluster Candidate X-ray Data: Physical
Quantities}
\tablewidth{0pt}
\tablehead{
\colhead{PDCS \#\tablenotemark{a}} & \colhead{Flux\tablenotemark{b}} &
\colhead{Luminosity\tablenotemark{b}} \\ 
\colhead{} & \colhead{($10^{-14}$ erg s$^{-1}$ cm$^{-2}$)} &
\colhead{($10^{43}$ erg s$^{-1}$)} \\ 
}
\startdata
57  & 5.7 & 1.5 \\ 
59  & 3.3 & 2.4 \\ 
60  & 6.4 & 0.3 \\ 
61  & 5.1 & 0.5 \\ 
62* &  21.5$^{+2.8}_{-2.6}$ & $ 5.6^{+0.7}_{-0.7}$  \\ 
63* &  8.7$^{+1.5}_{-1.4}$ & $ 5.4^{+0.9}_{-0.9}$  \\ 
64  & 3.2 & 4.4 \\ 
\enddata
\tablenotetext{a}{Clusters marked with a * are considered detections.}
\tablenotetext{b}{Errors are given only for detections, values without
errors are 3 $\sigma$ upper limits.}
\end{deluxetable}

\tablenum{6}
\begin{deluxetable}{lr}
\tablecaption{Fractional Change of Survey Volume from Model Variations.
\label{modelparams}}
\tablewidth{0pt}
\tablehead{
\colhead{Parameter Changed} & \colhead{Percent Change in
Volume} \\ 
}
\startdata
$q_o$ increased to 0.5 & -22\% \\
$3 \times r_c$ &  -11\% \\
$\frac{1}{3} \times r_c$ & +11\% \\
$\Delta\ M_{\star}$ = +0.1 & -9\% \\
$\Delta\ M_{\star}$ = -0.1 & +0.2\% \\
Sbc K corrections & +40\% \\
Poggianti K corrections & -10\% \\
\enddata

\end{deluxetable}

\tablenum{7}
\begin{deluxetable}{lrrr}
\tablecaption{Richness Function for ARC/KPNO and CFHT
Samples.\label{richdens} } 
\tablewidth{0pt}
\tablehead{
\colhead{Sample\tablenotemark{a}} & \colhead{Richness Class
1\tablenotemark{b}} & \colhead{Richness Class 2\tablenotemark{b}} &
\colhead{Richness Class 3\tablenotemark{b}} \\  
\colhead{} & \colhead{$40 \le \Lambda \le 60$} & \colhead{$60 \le
\Lambda \le 80$} &  \colhead{$80 \le \Lambda \le 120$} \\
\colhead{} & \colhead{$10^{-6}$ $h^3$ Mpc$^{-3}$} & \colhead{$10^{-6}$ $h^3$
Mpc$^{-3}$} & \colhead{$10^{-6}$ $h^3$ Mpc$^{-3}$} \\  
}
\startdata
ARC/KPNO -- V & $25.5^{+20.2}_{-12.2}$ (4) & $10.7^{+14.1}_{-6.9}$ (2) &
$4.9^{+11.3}_{-4.2}$ (1) \\
ARC/KPNO -- I & $12.7^{+16.8}_{-8.2}$ (2) & $10.7^{+14.1}_{-6.9}$ (2) & $4.9^{+11.3}_{-4.2}$ (1) \\
CFHT -- V      & $18.3^{+14.5}_{-8.8}$ (4) & $6.3^{+8.3}_{-4.1}$ (2) & $3.2^{+7.4}_{-2.8}$ (1) \\
CFHT -- I      & $9.1^{+7.2}_{-5.9}$ (2) & $6.3^{+8.3}_{-4.1}$ (2) & $3.2^{+7.4}_{-2.8}$ (1) \\
\enddata
\tablenotetext{a}{The four samples in this table are defined by the two
different selection processes from Sec. 4.2 and 4.3, and by using the $V$ and
$I$ band richnesses to determine the Richness Classes. }
\tablenotetext{b}{The value in parentheses is
the number of clusters from the sample in that Richness Class, as
defined by $\Lambda$.  The volumes are listed in Sec. 4.2 and 4.3.
We list the 68\% confidence limits based on a Poisson distribution
using the values from {Gehrels} 1986. 
}
\end{deluxetable}

\end{document}